\documentclass[12pt, hidelinks]{article}

\usepackage[T1]{fontenc}

\usepackage[utf8]{inputenc}
\usepackage[margin=1in]{geometry} 
\usepackage{amssymb,amsmath, physics}
\usepackage{pbox}
\usepackage{color, colortbl}
\usepackage{tensor}
\usepackage{wrapfig}
\usepackage{enumerate}
\usepackage{soul}
\usepackage{mathtools}
\usepackage{marvosym}
\usepackage{esint}
\usepackage{tikz}
\usepackage{stackengine}
\usepackage{subfigure}
\usepackage{color}
\definecolor{gg}{rgb}{0.0, 0.26, 0.15}
\usepackage[colorlinks=true,linkcolor=blue, citecolor=gg]{hyperref}
\usepackage{xcolor}

\newcommand{\hypgeo}[2]{%
  \operatorname{%
    {\vphantom{\mathnormal{F}}}_{#1}%
    \kern-\scriptspace
    \mathnormal{F}_{#2}%
  }%
}

\newcommand{\e}{\text{e}}

\newcommand{\eps}{\varepsilon}
\newcommand{\itr}[1]{%
  {\kern0pt#1}^{\mathrm{o}}%
}

\newcommand{\Sp}{\text{Sp}}

\begin{document}
\begin{center}
\LARGE
Ladder operators and coherent states for the Rosen-Morse system and its rational extensions\\
\vspace{1em}
\large
S. Garneau-Desroches$^\ast$, V. Hussin$^\dag$\\
\vspace{1em}
$^\ast$Département de Physique \& Centre de Recherches Mathématiques,\\
Université de Montréal, QC, H3C 3J7, Canada\\
$^\dag$Département de Mathématiques et de Statistique \& Centre de Recherches Mathématiques,\\
Université de Montréal, QC, H3C 3J7, Canada\\
\vspace{1em}
August 26, 2021
\end{center}

\begin{abstract}
Ladder operators for the hyperbolic Rosen-Morse (RMII) potential are realized using the shape invariance property appearing, in particular, using supersymmetric quantum mechanics. The extension of the ladder operators to a specific class of rational extensions of the RMII potential is presented and discussed. Coherent states are then constructed as almost eigenstates of the lowering operators. Some properties are analyzed and compared. 
The ladder operators and coherent states constructions presented are extended to the case of the trigonometric Rosen-Morse (RMI) potential using a point canonical transformation.
\end{abstract}

\section{Introduction}
Ladder operators are of interest in quantum mechanics due to their wide range of applications. Algebraic resolution of different systems \cite{Dirac.P.A.M, SS.40-02-12}, computation of matrix elements for observables \cite{SS.40-02-12}, study of underlying structure of quantum systems \cite{Dong.S-H} and coherent states construction \cite{CS.71} are common examples. For exactly solvable 1D quantum systems, ladder operators connect bounded eigenstates of adjacent energy levels of a Hamiltonian. Their realizations as first order differential operators in the case of an energy spectrum that is polynomial in the excitation number $n$ has been extensively studied (see, e.g., \cite{ Dong.S-H, LO.95-02-20, LO.08-05-05, LO.11-03-08}). This is however not so easy for systems where the energy is a rational function of the excitation number $n$ such as the hyperbolic Rosen-Morse (RMII) potential. This system admits a finite discrete bounded spectrum and has been studied in different contexts \cite{SS.32-10, SS.51-01-01, SS.09-04-22, SS.92-02-13, SY.12-10-26, SY.20-02-25}. Recently, a realization of the ladder operators was obtained for this system through an analogy with classical mechanics \cite{LO.20-06-18, LO.19-06}. In this paper, we intend to motivate a realization of the ladder operators for the RMII system from a purely quantum mechanical point of view.

Supersymmetric quantum mechanics (SUSYQM) was introduced in 1981 in the context of high-energy physics \cite{SY.81-04-29}. However, its application to non-relativistic quantum mechanics proliferated in the last decades since it can be used as a technique to generate new exactly solvable potentials from known initial ones with quasi-identical spectra. The hyperbolic Rosen-Morse system has been studied in the context of SUSYQM at the first order \cite{SY.12-10-26, Cooper.F} and the second order \cite{SY.20-02-25}. In particular, it is known that this potential is shape invariant \cite{Cooper.F, SY.83-09-25}, meaning that it returns to itself, with modified parameters, after a particular SUSY transformation. The aim of this work is to use this shape invariance to construct the ladder operators of the RMII system. We then make use of SUSYQM one step further in adapting the ladder operators to suit specific rational extensions (types of solvable potentials obtained from the RMII by a SUSY transformation), namely the type III from the classification \cite{SY.12-10-26}. 

Barut-Girardello coherent states of an infinite discrete bounded spectrum system are defined as eigenstates of a lowering ladder operator \cite{CS.71}. This definition as been extended for systems with a finite bounded spectrum \cite{CS.12-10-12, CS.08-07-15} as almost eigenstate of a lowering operator. Therefore, the realization of ladder operators motivates a precise coherent states construction which we work out both for the RMII system and for its rational extensions. Once built, standard coherent state properties such as space localization, trajectory of position and momentum expectation values, and minimization of the Heisenberg uncertainty principle are explored. 

It is also known that shape invariant potentials are connected through point canonical transformations (PCT) \cite{SS.92-02-13}. This offers a direct link to the trigonometric counterpart of the Rosen-Morse potential (RMI) \cite{tRM.05-12-21}, which also have energies rational in the excitation number $n$, but for which the discrete bounded spectrum is infinite. We exploit this connection to carry our results and constructions onto the RMI potential, therefore covering both Rosen-Morse potentials.

The plan of the paper is as follows. In Section \ref{sec:1}, we  review some of the known formalisms for ladder operators and first order supersymmetric quantum mechanics. In Section \ref{sec:2}, we introduce the hyperbolic Rosen-Morse (RMII) potential and expose its shape invariance property in SUSYQM. From there, the construction of ladder operators is carried out completely. In Section \ref{sec:3} we introduce the type III rational extensions of the RMII potential. We then proceed to adapt the RMII ladder operators to the type III rational extensions using SUSYQM. In Section \ref{sec:4}, we construct the associated coherent states as almost eigenstates of the lowering operators previously obtained for the RMII potential and its rational extensions. The space-localization, the trajectory and the position-momentum uncertainty relation are analyzed and then compared for these states. In Section \ref{sec:5}, we generalize the results to the trigonometric Rosen-Morse (RMI) system by means of a PCT. Conclusions are drawn in Section \ref{sec:7}, where further investigations are also suggested.

\section{Ladder operators and SUSYQM}\label{sec:1}
We present the theory on which the work is based. First, in Section \ref{sec:1.1}, we introduce ladder operators for 1D solvable quantum systems as operators connecting eigenstates of adjacent energies according to a precise action. A realization of this action is precisely what will be attempted for the Rosen-Morse systems, and will play an fundamental role in the coherent states construction. Then, a review of first order supersymmetric quantum mechanics (SUSYQM) is presented in Section \ref{sec:1.2_SUSYQM}. SUSYQM is a tool to generate new solvable systems from a known initial one. This formalism will play a role throughout this work both as a tool to realize the desired ladder operator action and as a way to extend the ladder operators to the rational extensions of the Rosen-Morse system.

\subsection{Ladder operators}\label{sec:1.1}
Consider a 1D exactly solvable quantum system described by a Hamiltonian $H$ and its associated Schrödinger equation ($ \hbar = 2m =1$):
\begin{align}\label{eq:1.1_SE}
H\psi(n;x) = E(n) \psi(n;x), \qquad \qquad H = -\dv[2]{x} + V(x), \qquad \qquad x \in \mathcal{D},
\end{align}
where $V(x)$ is the potential and $\psi(n;x) \in L^2(\mathcal{D})$ is a bounded eigenstate of $H$ with energy $E(n)$. The associated bounded spectrum $\Sp(H) = \qty{E(n)}_{n \geq 0}$  is discrete and can be either finite or infinite.

In this context, we usually define ladder operators $A^\pm$ as differential operators connecting eigenstates of $H$ of adjacent energies in $\Sp(H)$. Formally, we define $A^\pm$ by their action on the eigenstates \cite{LO.95-02-20, LO.08-05-05, LO.11-03-08}:
\begin{align}\label{eq:1.1_LO}
A^- \psi(n;x) = \sqrt{k(n)}\ \psi(n-1;x), \qquad \qquad A^+ \psi(n;x) = \sqrt{k(n+1)}\ \psi(n+1;x),
\end{align}
for a certain choice of a real positive function $k(n)$, and such that we have the ground state annihilation $A^- \psi(0;x) = 0$. In the case where the spectrum is finite, namely that there is a maximal excitation $n_{max}$, one may choose $k(n)$ to impose as well $A^+ \psi(n_{max};x) = 0$ (see, e.g., \cite{LO.08-02-25}). In this paper, we do not impose such restriction but consider the action $A^+ \psi(n_{max};x)$ to be ill-defined in the sense that it yields an unbounded state.

Introducing an operator $\delta(H)$ diagonal in the eigenstates basis, the operator $A^+ A^-$ factorizes the Hamiltonian as:
\begin{align}
A^+ A^- \psi(n;x) = \delta(H) \psi(n;x), \qquad \qquad \delta(E(n)) = k(n),
\end{align}
and similarly for the reverse product $A^- A^+$. Moreover, it is known \cite{LO.11-03-08, LO.20-06-18} that from the ladder operator action (\ref{eq:1.1_LO}), one obtains commutation relations with the Hamiltonian by introducing further diagonal operators $\Delta_{\pm}(H)$ and $\Omega(H)$:
\begin{align}
[H, A^\pm] \psi(n;x) &= A^\pm \Delta_\pm(H) \psi(n;x), \qquad \qquad \Delta_\pm(E(n)) = E(n \pm 1) - E(n),\\
[A^-, A^+] \psi(n;x) &= \Omega(H) \psi(n;x) , \qquad \qquad \qquad \Omega(E(n)) = k(n+1) - k(n).
\end{align}
From the later, one extracts the Generalized Heisenberg Algebra (GHA) \cite{LO.08-05-05, LO.11-03-08} generated by $\qty{H, A^-, A^+}$:
\begin{align}
[H,A^\pm] &= A^\pm \Delta_\pm(H), \qquad \qquad [A^-, A^+] = \Omega(H).
\end{align}

For common exactly sovable 1D quantum systems (harmonic oscillator, infinite square well, Morse, Pöschl-Teller, etc.), the realization (\ref{eq:1.1_LO}) for $A^\pm$ can be achieved with first order differential operators of the form \cite{Dong.S-H}:
\begin{align}\label{eq:1.1_O1_LO}
A^\pm = f^\pm(x,N) + g^\pm(x,N) \dv{x},
\end{align}
where $f^\pm$ and $g^\pm$ are to be determined and where $N$ is the diagonal number operator defined by:
\begin{align}
N \psi(n;x) = n \psi(n;x).
\end{align}
The use of the number operator in the construction of $A^\pm$ arise from the fact that the operators needs to vary depending on the excitation of the state it acts on. Standard construction techniques can be found in \cite{Dong.S-H, LO.01-10-19, LO.01-12-18, LO.14-06}. However, in this work we consider the Rosen-Morse system, for which a first order realization such as (\ref{eq:1.1_O1_LO}) cannot be achieved using standard methods \cite{LO.20-06-18}.

\subsection{Supersymmetric quantum mechanics}\label{sec:1.2_SUSYQM}
Suppose two Hamiltonians $H$ and $\tilde{H}$ are connected by intertwining operators $B^\pm$ in the following way \cite{Cooper.F, SY.85-12.2917, SY.10-10-18, SY.04-12-21, SY.84-06-15}:
\begin{align}\label{eq:1.2_intertwining_rel}
B^- H = \tilde{H} B^-, \qquad \qquad H B^+ = B^+ \tilde{H}.
\end{align}
For first order SUSYQM, we take $B^\pm$ to be differential operators of the first order:
\begin{align}\label{eq:1.2_intertwining_op}
B^\pm = W(x) \pm \dv{x},
\end{align}
where the superpotential $W(x)$ is a real function. Necessary and sufficient conditions for (\ref{eq:1.2_intertwining_rel}) are then obtained:
\begin{align}
\tilde{V}(x) = V(x) -2 W'(x), \label{eq:1.2_partner_pot}\\
 W'(x) + W^2(x) = V(x) - \eps \label{eq:1.2_ricatti}, 
\end{align}
where $\eps \in \mathbb{R}$ is an integration constant referred to as the factorization energy. We know that SUSYQM is a technique to generate new exactly solvable systems from a known initial one \cite{SY.10-10-18}. Taking $H$ to be such initial solved system, it is seen that finding a couple $(W(x), \eps)$ solving the Ricatti equation (\ref{eq:1.2_ricatti}) yields an expression for the new potential $\tilde{V}(x)$ through (\ref{eq:1.2_partner_pot}). Solutions are usually achieved by setting $W(x) = u'(x) / u(x)$. Equation (\ref{eq:1.2_ricatti}) thus reduces to a Schrödinger equation for $u(x)$ with energy $\eps$:
\begin{align}\label{eq:1.2_seed_SE}
-u''(x) + V(x) u(x) = \eps u(x).
\end{align}
The seed solution $u(x)$ need not be normalizable, but has to be nodeless in order to avoid singularities in $W(x)$. The factorization energy is restricted by $\eps \leq E(0)$ accordingly.

It is well known that the intertwining relations (\ref{eq:1.2_intertwining_rel}) allow to obtain the eigenstates of the new system $\tilde{H}$ from that of the initial one $H$. Indeed, we have:
\begin{align}
\tilde{H}(B^- \psi(n;x)) = B^- H \psi(n;x) = E(n) (B^- \psi(n;x)),
\end{align}
making $B^- \psi(n;x)$ an eigenstate of $\tilde{H}$ with energy $E(n)$ unless $B^- $ annihilates $\psi(n;x)$. Normalized eigenstates $\tilde{\psi}(n;x)$ and energies $\tilde{E}(n)$ are recovered: 
\begin{align}\label{eq:1.2_connection}
\tilde{\psi}(n;x) = \frac{B^- \psi(m;x)}{\sqrt{E(m) - \eps}}, \qquad \qquad \psi(m;x) = \frac{B^+ \tilde{\psi}(n;x)}{\sqrt{\tilde{E}(n) - \eps}}, \qquad \qquad \tilde{E}(n) = E(m),
\end{align}
where $m = n \pm 1$ according to whether an energy level is created or suppressed during the transformation. If the spectra agree perfectly, we have $m =n$ instead. Let us shortly exhibit the three cases for further referring. More details can be found in \cite{SY.04-12-21, SY.03-02-26, Kuru.S, LO.19-05-23}.

\begin{enumerate}
\item[(a)] State-deleting SUSY ($m = n+1$). This case arises when $u(x)$ is the ground state $\psi(0;x)$ of $H$ with $\eps = E(0)$. Every state of $H$ is connected to one state of $\tilde{H}$ according to (\ref{eq:1.2_connection}), except for the ground state since $B^- \psi(0;x) = 0$. The energy level $E(0)$ is removed of the spectrum of $\tilde{H}$ during the transformation: $\Sp(\tilde{H}) = \Sp(H) \setminus \qty{E(0)}$. The ground state of $\tilde{H}$ has energy $\tilde{E}(0) = E(1)$.

\item[(b)] State-adding SUSY ($m = n-1$). This case arises for unbounded seed solutions $u(x)$ with $\eps < E(0)$ such that $1/u(x)$ is normalizable. All the states of $H$ are connected to that of $\tilde{H}$ according to (\ref{eq:1.2_connection}). It is known however in this case that $B^+$ annihilates $1/u(x)$, making it a normalizable eigenstate of $\tilde{H}$ with energy $\eps < E(0)$. An energy level $\eps$ is created during the transformation: $\Sp(\tilde{H}) = \Sp(H) \cup \qty{\eps}$. The ground state of the new system is thus:
\begin{align}
\tilde{\psi}(0;x) \propto \frac{1}{u(x)}, \qquad \qquad \tilde{E}(0) = \eps.
\end{align}

\item[(c)] Isospectral SUSY ($m=n$). This case arises when both $u(x)$ and $1/u(x)$ are unbounded. Every states and energy levels of $H$ are in exact correspondence with that of $\tilde{H}$ without any suppression nor creation of energy levels during the transformation: $\Sp(\tilde{H}) = \Sp(H)$.
\end{enumerate}
Finally, the operators $B^\pm$ allow the factorization of the Hamiltonians \cite{SY.04-12-21}:
\begin{align}\label{eq:1.2_facto}
H = B^+B^- + \eps, \qquad \qquad \tilde{H} = B^-B^+ + \eps.
\end{align}

\section{RMII ladder operators from SUSYQM}\label{sec:2}
We first study the hyperbolic Rosen-Morse (RMII) system which was introduced in 1932 as an exactly solvable quantum system of use in the modelization of vibrations in polyatomic molecules \cite{SS.32-10}. The known solutions and the associated spectrum is presented in Section \ref{sec:2.1_RMII}. The potential is given by two parameters: $\lambda$ dictates the value on the boundaries of the domain, while $s$ controls the attraction (depth) of the potential. In Section \ref{sec:2.2_SI_RMII}, we expose how a change in the parameter $s$, known as shape invariance, occurs when the RMII potential goes under a state-deleting SUSY transformation. For this reason, we will label the RMII potential by $V_s(x)$ according to the value of $s$, while taking $\lambda$ to be fixed. This shape invariance property generates a hierarchy of RMII potentials for different values of $s$. In Section \ref{sec:2.3_Construction_LO}, we show how the connection between members of the hierarchy using SUSYQM allows the construction of ladder operators $A^\pm$ for a specific (fixed $s$) member. This new ladder operators realization is presented step by step.

\subsection{The RMII potential} \label{sec:2.1_RMII}

\begin{figure}[h]
\centering
\includegraphics[width=0.40\textwidth]{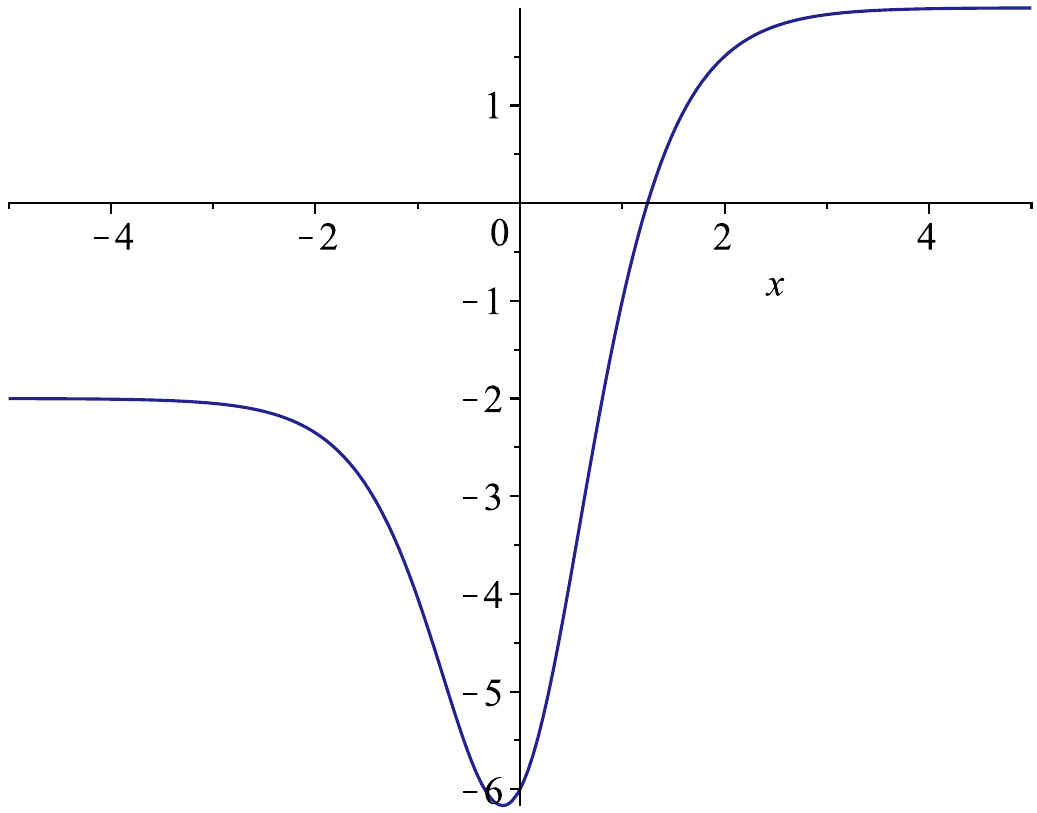}
\caption{Hyperbolic Rosen-Morse (RMII) potential $V_s(x)$ with parameters $s = 2$ and $\lambda = 1$.}
\label{fig:2.1_RMII}
\end{figure}

The hyperbolic Rosen-Morse (RMII) potential is defined as \cite{SS.32-10, SS.09-04-22, SY.20-02-25}:
\begin{align}\label{eq:2.1_RMII}
V_s(x) = 2 \lambda \tanh{x} - s(s+1) \sech^2{x}, \qquad \qquad x \in \mathbb{R}.  
\end{align}
We can without loss of generality assume $\lambda \geq 0$. We further impose $s > 0$ and $0 \leq \lambda < s^2$, ensuring the well-shape of the potential and the existence of at least one bounded eigenstate. Solutions to the Schrödinger equation (\ref{eq:1.1_SE}) allows for a finite number of bounded sates and for scattering states due to the asymptotic behaviour:
\begin{align}\label{eq:2.1_asymptote}
\lim_{x \to \pm \infty} V_s(x) = \pm 2 \lambda,
\end{align}
as shown in Figure \ref{fig:2.1_RMII}. 

This work focuses on the normalizable eigenstates which are expressed in terms of the Jacobi polynomials $P_{n}^{(\alpha, \beta)}(y)$ \cite{Temme.N} as:
\begin{align}
\psi_s(n;x) = M_s(n) \cosh^{-(s-n)}(x)\ \e^{-\frac{\lambda x}{s-n}}\  P_n^{(a_s(n),b_s(n))}(\tanh{x}), \label{eq:2.1_e.state}
\end{align}
for which the parameters take the form:
\begin{align}
a_s(n) = s-n + \frac{\lambda}{s-n}, \qquad \qquad b_s(n) = s-n - \frac{\lambda}{s-n}.
\end{align}
Here, $M_s(n)$ is the normalization constant given by \cite{SS.09-04-22, SS.78-04-01}:
\begin{align}
M_s(n) = (-1)^n 2^{n-s} \sqrt{\frac{n!  \qty((s-n)^2-\frac{\lambda^2}{(s-n)^2})  \Gamma(2s-n+1)}{(s-n)  \Gamma\qty(s+1+\frac{\lambda}{s-n}) \Gamma\qty(s+1-\frac{\lambda}{s-n})}}.
\end{align}
The associated bounded spectrum is finite and the energies are rational in the excitation number $n$:
\begin{align}
E_s(n) = -(s-n)^2- \frac{\lambda^2}{(s-n)^2}, \qquad \qquad n = 0,1,\dots, n_{max} < s- \sqrt{\lambda},
\end{align}
where the upper bound on $n_{max}$ prevents the energy to exceed the lowest asymptote in (\ref{eq:2.1_asymptote}).

\subsection{SUSYQM and shape invariance for the RMII potential} \label{sec:2.2_SI_RMII}
We perform a first order state-deleting SUSY transformation on the RMII potential using the ground state as seed solution:
\begin{align}\label{eq:2.3_seed}
u(x) = \psi_s(0;x) = M_s(0)\ \cosh^{-s}(x)\ \e^{-\frac{\lambda x}{s}} , \qquad \qquad \eps = E_s(0) = -s^2 - \frac{\lambda^2}{s^2}.
\end{align}
Following the steps from Section \ref{sec:1.2_SUSYQM}, the intertwining operators and the superpotential are:
\begin{align}
B^\pm_s = W_s(x) \pm \dv{x}, \qquad \qquad W_s(x) = -s \tanh{x} - \frac{\lambda}{s}.
\end{align}
Then, the partner potential is obtained from equation (\ref{eq:1.2_partner_pot}) and takes the form:
\begin{align}
\tilde{V}_s(x) = 2 \lambda \tanh{x} - s(s-1)\sech^2{x},
\end{align}
where $\tilde{V}_s(x)$ is in fact a RMII potential with translated parameter $s \to s-1$. The eigenstates connection (\ref{eq:1.2_connection}) of the state-deleting SUSY is thus established to be:
\begin{align}\label{eq:2.3_e.state_connection}
\psi_{s-1}(n;x) = \frac{B_s^- \psi_s(n+1;x)}{\sqrt{E_s(n+1) - E_s(0)}}, \qquad \qquad \psi_{s}(n+1;x) = \frac{B_s^+ \psi_{s-1}(n;x)}{\sqrt{E_{s-1}(n) - E_s(0)}},
\end{align}
with the energy relation:
\begin{align}\label{eq:2.3_energy_connection}
E_{s-1}(n) = E_{s}(n+1), \qquad  \qquad n =0,1,2,\dots, n_{max} -1 .
\end{align}

\begin{figure}[h]
\centering
\begin{tikzpicture}

\node at (-0.9,1) {\scriptsize $\psi_s(0)$};
\node at (-0.9,1.6) {\scriptsize $\psi_s(1)$};
\node at (-0.9,2.7) {\scriptsize $\psi_s(n-1)$};
\node at (-0.9,3.3) {\scriptsize $\psi_s(n)$};

\node at (-1.1,4.1) {{$H_s$}};
\draw (-1.5,0.8) -- (-0.3,0.8);
\draw (-1.5,1.4) -- (-0.3,1.4);
\node at (-0.9, 2.2) {$\vdots$};
\draw (-1.5,2.5) -- (-0.3,2.5);
\draw (-1.5,3.1) -- (-0.3,3.1);

\draw[gg, dotted] (-2.2,0.8) -- (-1.7,0.8);
\draw[gg, dotted] (-2.2,1.4) -- (-1.7,1.4);
\node at (-2.5, 2.2) {$\vdots$};
\draw[gg, dotted] (-2.2,2.5) -- (-1.7,2.5);
\draw[gg, dotted] (-2.2,3.1) -- (-1.7,3.1);

\node[gg] at (-2.7,0.8) {\scriptsize $E_s(0)$};
\node[gg] at (-2.7,1.4) {\scriptsize $E_s(1)$};
\node[gg] at (-3,2.5) {\scriptsize $E_s(n-1)$};
\node[gg] at (-2.7,3.1) {\scriptsize $E_s(n)$};


\node at (1.8,1.6) {\scriptsize $\psi_{s-1}(0)$};
\node at (1.8,2.7) {\scriptsize $\psi_{s-1}(n-2)$};
\node at (1.8,3.3) {\scriptsize $\psi_{s-1}(n-1)$};


\node at (1.8,4.1) {{$H_{s-1}$}};
\draw (1.2,1.4) -- (2.4,1.4);
\node at (1.8, 2.2) {$\vdots$};
\draw (1.2,2.5) -- (2.4,2.5);
\draw (1.2,3.1) -- (2.4,3.1);


\node at (7.2,4.1) {{$H_{s-n+1}$}};
\draw (6.6,2.5) -- (7.8,2.5);
\draw (6.6,3.1) -- (7.8,3.1);
\node at (7.2,2.7) {\scriptsize $\psi_{s-n+1}(0)$};
\node at (7.2,3.3) {\scriptsize $\psi_{s-n+1}(1)$};

\node at (9.9,4.1) {{$H_{s-n}$}};
\draw (9.3,3.1) -- (10.5,3.1);
\node at (9.9,3.3) {\scriptsize $\psi_{s-n}(0)$};

\node at (4.5,2.7) {$\dots$};
\node at (4.5,3.1) {$\dots$};

\draw[red,->] (9,3.5) .. controls (8.6,3.7) .. (8.2,3.5);
\draw[red,->] (6.2,3.5) .. controls (5.6,3.7) .. (5,3.5);
\draw[red,->] (4,3.5) .. controls (3.4,3.7) .. (2.8,3.5);
\draw[red,->] (0.8,3.5) .. controls (0.4,3.7) .. (0,3.5);

\draw[red,->]  (8.2,3) .. controls (8.6,2.8) ..(9,3) ;
\draw[red,->] (5,3) .. controls (5.6,2.8) .. (6.2,3);
\draw[red,->] (2.8,3) .. controls (3.4,2.8) .. (4,3);
\draw[red,->]  (0,3) .. controls (0.4,2.8) ..(0.8,3) ;

\node[red] at (8.6,3.9) {\scriptsize $B^+_{s-n+1}$};
\node[red] at (5.6,3.9) { \scriptsize $B^+_{s-n+2}$};
\node[red] at (3.4,3.9) {\scriptsize $B^+_{s-1}$};
\node[red] at (0.5,3.9) {\scriptsize $B^+_s$};

\node[red] at (8.6,2.6) {\scriptsize $B^-_{s-n+1}$};
\node[red] at (5.6,2.6) { \scriptsize $B^-_{s-n+2}$};
\node[red] at (3.4,2.6) {\scriptsize $B^-_{s-1}$};
\node[red] at (0.5,2.6) {\scriptsize $B^-_s$};

\end{tikzpicture}
\caption{Connection between the $n$-th excited state $\psi_s(n;x)$ of $H_s$ and the ground state $\psi_{s-n}(0;x)$ of $H_{s-n}$ with operators $B^\pm_s$.}
\label{fig:2.3_hierarchy}
\end{figure}
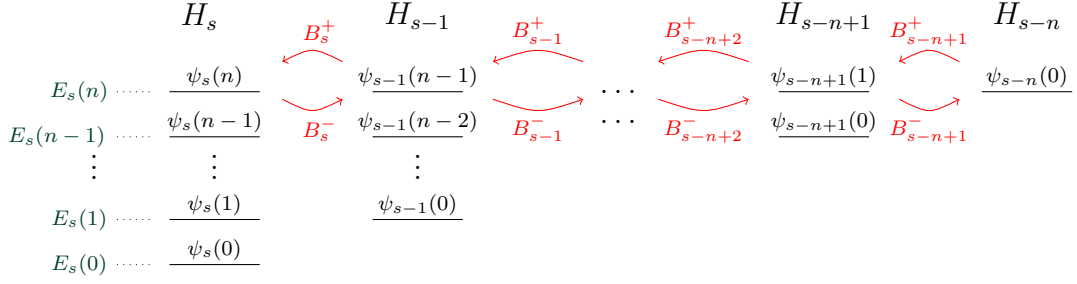

Indeed, the relation $\tilde{V}_{s}(x) = V_{s-1}(x)$ is known as shape invariance \cite{Cooper.F, SY.85-12.L57, SY.87-04-06} and allows the following hierarchy generating procedure. One could now take $\tilde{V}_s = V_{s-1}$ as the starting potential for that same state-deleting SUSY transformation and obtain another RMII partner potential $\tilde{V}_{s-1} = V_{s-2}$, invoquing the shape invariance a second time. Thus, performing this SUSY transformation iteratively by translating the $s$ parameter accordingly at each step generates a hierarchy of RMII potentials $\qty{V_s, V_{s-1}, \dots, V_{s-n_{max}}}$.  The eigenstates connection of any two adjacent Hamiltonians in the hierarchy is ensured by the appropriate $B^\pm_s$ operators following (\ref{eq:2.3_e.state_connection}). Moreover, loosing the ground state energy level at each stage of the iteration, it becomes possible to express any eigenstate $\psi_s(n;x)$ of $H_s$ in terms of the ground state of another Hamiltonian in the hierarchy \cite{Cooper.F}. Precisely, we have:
\begin{align}
\psi_s(n;x) = \frac{B^+_s}{\sqrt{E_s(n)-E_s(0)}} \frac{B^+_{s-1}}{\sqrt{E_s(n)-E_s(1)}} \cdots \frac{B^+_{s-n+1}}{\sqrt{E_s(n)-E_s(n - 1)}} \psi_{s-n}(0;x), \label{eq:2.3_up_connection}
\end{align}
where we have used the energy relation (\ref{eq:2.3_energy_connection}) to express all the energies appearing in the formula as energies of the $H_s$ system. It is through this precise connection (\ref{eq:2.3_up_connection}) that the equivalence between shape invariance in SUSYQM and the Factorization Method of Infeld and Hull \cite{SS.51-01-01} is establisehd \cite{SY.89-04}. The relation converse to (\ref{eq:2.3_up_connection})  is obtained with the $B^-_s$:
\begin{align}
\psi_{s-n}(0;x) = \frac{B^-_{s-n+1}}{\sqrt{E_s(n)-E_s(n-1)}} \frac{B^-_{s-n +2 }}{\sqrt{E_s(n)-E_s(n-2)}} \cdots \frac{B^-_{s}}{\sqrt{E_s(n)-E_s(0)}} \psi_{s}(n;x). \label{eq:2.3_down_connection}
\end{align}
The connections (\ref{eq:2.3_up_connection}) and  (\ref{eq:2.3_down_connection}) in the hierarchy are illustrated in Figure \ref{fig:2.3_hierarchy}, where each column fixes a specific system while each row fixes a specific value of energy. Indeed, action with $B^\pm_s$ does not affect the value of the energy, but modifies the place this energy occupies in the spectrum.

Hence, shape invariance allows the connection between eigenstates of different RMII systems having the same energy. These horizontal displacements in Figure \ref{fig:2.3_hierarchy} allowed by the $B^\pm_s$ will play a central role in the construction of the RMII ladder operators.

\subsection{Construction of the ladder operators}\label{sec:2.3_Construction_LO}
The previous section highlighted our ability to link eigenstates of different RMII systems for fixed energy. What remains is to find a way to connect at least two eigenstates of adjacent energy levels, regardless of the systems to which they belong in the hierarchy. This idea has been explored algebraically in \cite{CS.93-09-13}. Once this is achieved, we can combine the different actions to construct ladder operators that connects eigenstates of adjacent energy levels within a fixed system.

\begin{figure}[h]
\centering
\begin{tikzpicture}

\node at (-0.9,1) {\scriptsize $\psi_s(0)$};
\node at (-0.9,1.6) {\scriptsize $\psi_s(1)$};
\node at (-0.9,2.7) {\scriptsize $\psi_s(n-1)$};
\node at (-0.9,3.3) {\scriptsize $\psi_s(n)$};

\node at (-1.1,4.1) {{$H_s$}};
\draw (-1.5,0.8) -- (-0.3,0.8);
\draw (-1.5,1.4) -- (-0.3,1.4);
\node at (-0.9, 2.2) {$\vdots$};
\draw (-1.5,2.5) -- (-0.3,2.5);
\draw (-1.5,3.1) -- (-0.3,3.1);

\draw[gg, dotted] (-2.2,0.8) -- (-1.7,0.8);
\draw[gg, dotted] (-2.2,1.4) -- (-1.7,1.4);
\node at (-2.5, 2.2) {$\vdots$};
\draw[gg, dotted] (-2.2,2.5) -- (-1.7,2.5);
\draw[gg, dotted] (-2.2,3.1) -- (-1.7,3.1);

\node[gg] at (-2.7,0.8) {\scriptsize $E_s(0)$};
\node[gg] at (-2.7,1.4) {\scriptsize $E_s(1)$};
\node[gg] at (-3,2.5) {\scriptsize $E_s(n-1)$};
\node[gg] at (-2.7,3.1) {\scriptsize $E_s(n)$};


\node at (1.8,1.6) {\scriptsize $\psi_{s-1}(0)$};
\node at (1.8,2.7) {\scriptsize $\psi_{s-1}(n-2)$};
\node at (1.8,3.3) {\scriptsize $\psi_{s-1}(n-1)$};


\node at (1.8,4.1) {{$H_{s-1}$}};
\draw (1.2,1.4) -- (2.4,1.4);
\node at (1.8, 2.2) {$\vdots$};
\draw (1.2,2.5) -- (2.4,2.5);
\draw (1.2,3.1) -- (2.4,3.1);


\node at (7.2,4.1) {{$H_{s-n+1}$}};
\draw (6.6,2.5) -- (7.8,2.5);
\draw (6.6,3.1) -- (7.8,3.1);
\node at (7.2,2.7) {\scriptsize $\psi_{s-n+1}(0)$};
\node at (7.2,3.3) {\scriptsize $\psi_{s-n+1}(1)$};

\node at (10.1,4.1) {{$H_{s-n}$}};
\draw (9.5,3.1) -- (10.5,3.1);
\node at (10.1,3.3) {\scriptsize $\psi_{s-n}(0)$};

\node at (4.5,2.7) {$\dots$};
\node at (4.5,3.1) {$\dots$};

\node[red] at (0.5,1.7) {\scriptsize $\gamma^{-1}_{s}(x)$};
\node[red] at (0.9,0.6) {\scriptsize $\gamma_{s}(x)$};

\draw[red, ->] (0.9, 1.5) .. controls (0.5,1.5) .. (0.1,1.1);
\draw[red, ->] (0.2,0.8).. controls (0.8,0.8) ..  (1, 1.2);

\node[red] at (8.7,3.4) {\scriptsize $\gamma^{-1}_{s-n+1}(x)$};
\node[red] at (9.0,2.3) {\scriptsize $\gamma_{s-n+1}(x)$};

\draw[red, ->] (9.0, 3.2) .. controls (8.5,3.2) .. (8.1,2.8);
\draw[red, ->] (8.2,2.5).. controls (8.8,2.5) ..  (9.2, 2.9);

\node[red] at (3.2,2.2) {\scriptsize $\gamma^{-1}_{s-1}(x)$};
\node[red] at (3.2,1.1) {\scriptsize $\gamma_{s-1}(x)$};

\draw[red, ->] (3.7, 2.0) .. controls (3.2,2.0) .. (2.8,1.6);
\draw[red, ->] (2.9,1.3).. controls (3.5,1.3) ..  (3.9, 1.7);

\node[red] at (5.6,2.8) {\scriptsize $\gamma^{-1}_{s-n+2}(x)$};
\node[red] at (5.6,1.7) {\scriptsize $\gamma_{s-n+2}(x)$};

\draw[red, ->] (6.1, 2.6) .. controls (5.6,2.6) .. (5.2,2.2);
\draw[red, ->] (5.3,1.9).. controls (5.9,1.9) ..  (6.3, 2.3);

\node at (4.5,1.8) {$\dots$};
\node at (4.5,2.2) {$\vdots$};
\end{tikzpicture}
\caption{Connection between ground states of adjacent members of the RMII hierarchy by action with $\gamma_s(x)$ or $\gamma^{-1}_s(x)$. }
\label{fig:2.4_GS_connection}
\end{figure}
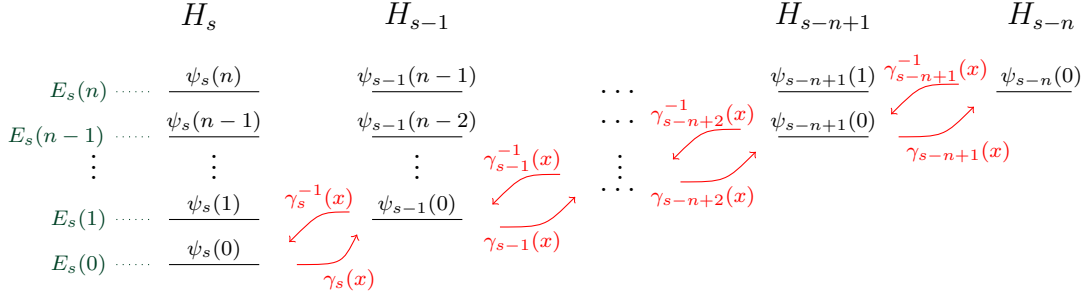

In our case, we establish this connection for different energies  between the ground states (\ref{eq:2.3_seed}) of adjacent members of the hierarchy. Defining:
\begin{align}
\gamma_s(x) = \cosh(x)\ \e^{-\frac{\lambda x}{s(s-1)}}, 
\end{align}
we have the relations
\begin{align}
\psi_{s-1}(0;x) = \qty(\frac{M_{s-1}(0)}{M_s(0)} \gamma_s(x)) \psi_s(0;x), \label{eq:2.4_GS_lower}
\end{align}
and
\begin{align}
\psi_{s+1}(0;x) = \qty(\frac{M_{s+1}(0)}{M_s(0)} \gamma^{-1}_{s+1}(x)) \psi_{s}(0;x). \label{eq:2.4_GS_raise}
\end{align}
As ground state energy is what is lost or gained as we switch from adjacent systems, the action of $\gamma_s(x)$ or its inverse respectively increase or decrease energy. Hence, equations (\ref{eq:2.4_GS_lower}) and (\ref{eq:2.4_GS_raise}) amount to vertical displacements in the hierarchy scheme, as illustrated in Figure \ref{fig:2.4_GS_connection}. 

Now composing the action of the $B^\pm_s, \gamma_s(x)$ and $\gamma^{-1}_s(x)$, we can finally construct the ladder operators for the RMII system $H_s$ for $s$ fixed. Starting with an arbitrary state $\psi_s(n;x)$, we first act with the appropriate product of the $B^-_s$ operators following equation (\ref{eq:2.3_down_connection}) to reach the ground state $\psi_{s-n}(0;x)$. At this point the energy has remained fixed in the process. From there, we act with $\gamma_{s-n}(x)$ or $\gamma^{-1}_{s-n+1}(x)$ respectively for a raising or lowering action. The energy has been raised or lowered accordingly during this step. Finally, we go back to the $H_s$ system by acting successively with the appropriate products of $B^+_s$ operators following (\ref{eq:2.3_up_connection}). The states obtained are respectively $\psi_s(n+1;x)$ in the case of a raising action and $\psi_s(n-1;x)$ for a lowering action. While the initial product of $B^-_s$ only depends on $n$, the final product of $B^+_s$ depends also on the raising or lowering nature of the action. Taking the constants into account, the expression for the ladder operators $A^\pm(n)$ acting on the $n$-th excited eigenstate $\psi_s(n;x)$ of $H_s$ are given as ordered products:
\begin{align}
 A^+(n) &  = B_s^+ \qty(\prod_{i=1}^n \frac{B^+_{s-i}}{\sqrt{E_s(n+1)-E_s(i)}}) \qty( \frac{M_{s-n-1}(0)}{M_{s-n}(0)} \gamma_{s-n}(x)) \qty(\prod_{i=1}^n \frac{B^-_{s-n+i}}{\sqrt{E_s(n) - E_s(n-i)}}), \label{eq:2.4_raising}\\
A^-(n) & = \qty(\prod_{i=0}^{n-2} \frac{B^+_{s-i}}{\sqrt{E_s(n-1)-E_s(i)}}) \qty(\frac{M_{s-n+1}(0)}{M_{s-n}(0)} \gamma^{-1}_{s-n+1}(x)) \qty(\prod_{i=1}^{n-1} \frac{B^-_{s-n+i}}{\sqrt{E_s(n) - E_s(n-i)}}) B^-_s. \label{eq:2.4_lowering}
\end{align}
The obtained ladder operators $A^\pm(n)$ are differential operators in $x$ of order $2n \pm 1$ except for $A^-(0)$ which is also of the first order. This family of operators realize the ladder operator action (\ref{eq:1.1_LO}):
\begin{align}
A^-(n)\ \psi_s(n;x) = \sqrt{k(n)}\ \psi_s(n-1;x), \quad \quad A^+(n)\ \psi_s(n;x) = \sqrt{k(n+1)}\ \psi_{s}(n+1;x),
\end{align}
with $k(n)$ being the shifted energy:
\begin{align}\label{eq:2.4_shifted_energy}
k(n) = E_s(n) - E_s(0).
\end{align}
We insist on the fact that the choice of $k(n)$ to be the shifted energy is not only a common one \cite{CS.07-05-30, CS.12-10-12, CS.08-07-15}, but arises naturally from our construction as it appears in the normalization factor of every SUSY transformation (recall (\ref{eq:2.3_e.state_connection})). We further wish to stress the fact that we have used the different members (different $s$) of the hierarchy only at intermediate steps in the construction of the ladder operators for a precise (fixed $s$) RMII system $H_s$.

\section{Application to type III rational extensions}\label{sec:3}
In this section, we wish to extend our ladder operators realization to other SUSY partners of the RMII system. First order rational extensions are particular state-adding or isospectral SUSY partners arising from a polynomial seed solution \cite{SY.04-01-19, SY.13-12-03, SY.10-01-24, SY.12-02, SY.13-03-29, SY.20-04-28, SY.09-08-12}. The expression for these partner potentials can be decomposed in two terms: one of which is the initial potential with modified parameters and the other being a rational function of some variable. In the case of the RMII system, rational extensions have been classified into three distinct classes in \cite{SY.12-10-26}. The first two classes, referred to as "type I" and "type II" in \cite{SY.12-10-26}, arise from isospectral SUSY transformations and affect only slightly the shape of the initial potential. They are not considered in this work. On the other hand, the last class, "type III", arises as a state-adding SUSY transformation and drastically modify the shape of the initial potential to allow for the extra energy level $\eps$.  In Section \ref{sec:3.1}, we briefly present the type III rational extensions of the RMII potential. In Section \ref{sec:3.2_LO_RE}, we construct ladder operators for these partner systems using that of the RMII. Technicalities involving the additional energy level are addressed.

\subsection{Type III rational extensions of the RMII potential}\label{sec:3.1}
\begin{figure}[h]
    \centering
  \includegraphics[width=0.55\textwidth]{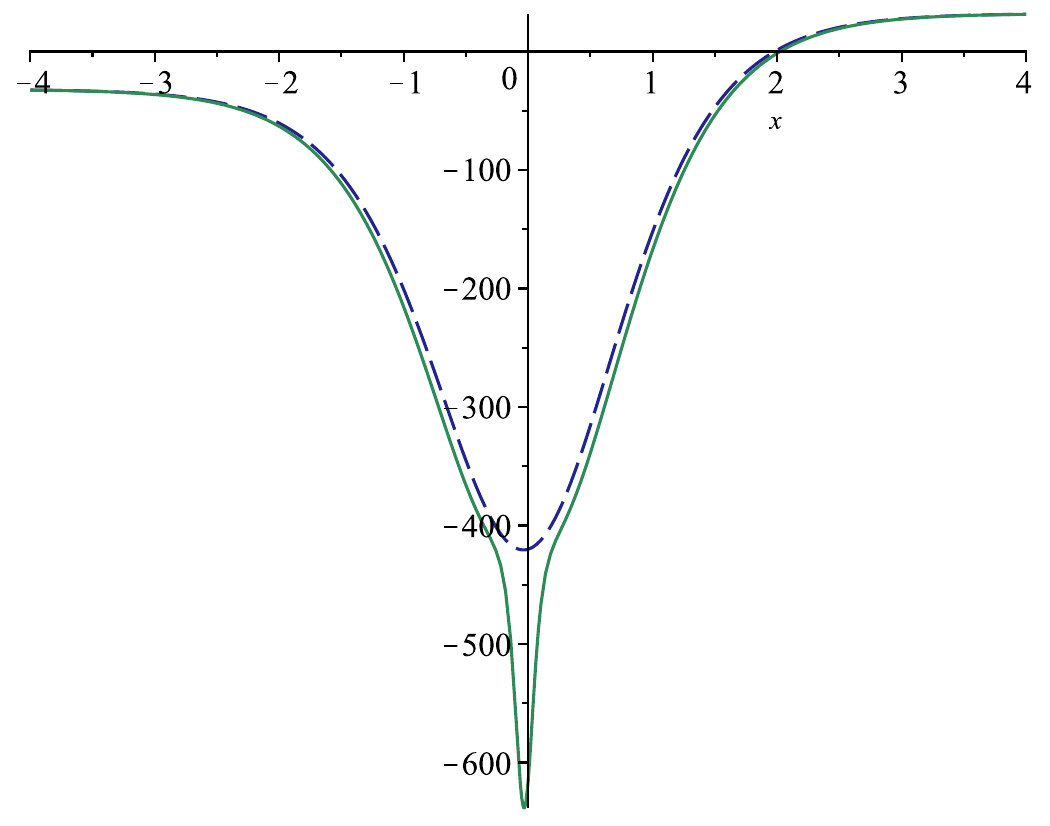}
    \caption{Initial RMII potential $V_s(x)$ (blue, dashed) and its type III rational extension SUSY partner potential $\tilde{V}(x)$ (green, solid) for parameters $s = 20$, $\lambda = 16$ and $k = 2$. The additional energy level is $\eps \approx -529.48$.}
\label{fig:3.1_graph_RE}
\end{figure}

The seed solutions used to generate a type III rational extension of RMII are nodeless unbounded polynomial solutions, in the variable $z = \tanh{x}$,  of the associated Schrödinger equation with factorization energy $\eps < E_s(0)$. These seed solutions are given by \cite{SY.12-10-26}:
\begin{align}
u(x) = (1-z)^{-\frac{\tilde{a}}{2}} (1+z)^{-\frac{\tilde{b}}{2}} P_{k}^{(-\tilde{a}, -\tilde{b})}(z), \qquad \eps = -(s+k+1)^2 - \frac{\lambda^2}{(s+k+1)^2},
\end{align}
where $k \in \qty{2,4,6,\dots}$  corresponds to the degree of the different polynomial solutions available\footnote{Taking polynomials of even degree ensures $u(x)$ to be nodeless for $z \in (-1, +1)$ (see \cite{SY.12-10-26}).}. Once chosen, $k$ remains fixed for the transformation. The associated parameters are:
\begin{align}
\tilde{a} = s+k+1 + \frac{\lambda}{s+k+1}, \qquad \qquad \tilde{b} = s+k+1 - \frac{\lambda}{s+k+1}.
\end{align}
It is shown that $1/u(x)$ is normalizable \cite{SY.12-10-26}, hence these polynomial seed solutions are indeed candidates for state-adding SUSY transformations. Fixing $k$ and implementing the SUSY transformation, the intertwining operators are:
\begin{align}
\mathcal{B}^\pm = (s+k+1) \tanh{x} + \frac{\lambda}{s+k+1} + \dv{x} \log P^{(-\tilde{a},-\tilde{b})}_k(\tanh{x}) \pm \dv{x}.
\end{align}
The three first terms correspond to the superpotential, from which the partner potential $\tilde{V}(x)$ is derived. After manipulations, the type III rational extensions of the RMII potential are \cite{SY.12-10-26}:
\begin{align}\label{eq:3.1_RE}
\tilde{V}(x) = V_{s+1}(x) + 2(1-z^2) \qty{2z \frac{P'_k(z)}{P_k(z)} -(1-z^2)\qty[\frac{P''_k(z)}{P_k(z)} - \qty(\frac{P'_k(z)}{P_k(z)})^2]-k},
\end{align}
where we have abbreviated $P_k(z) \equiv P^{(-\tilde{a},-\tilde{b})}_k(z)$ and where the prime denotes differentiation with respect to $z$. The first term is a RMII potential with parameter $s$ translated by $+1$, while the second term is rational in $z$. Figure \ref{fig:3.1_graph_RE} provides a comparison of the two partner potentials for this SUSY transformation. The well is dug from the RMII system to the rational extension in order to allow the additional energy level $\eps$.

The normalizable eigenstates for the rational extensions are obtained via the correspondence (\ref{eq:1.2_connection}) for state-adding SUSY:
\begin{align}
\begin{cases}
\tilde{\psi}(n+1;x) = \frac{\mathcal{B^-} \psi_s(n;x)}{\sqrt{E_s(n) - \eps}}, & \qquad \qquad  \tilde{E}(n+1) = E_s(n),\\
\tilde{\psi}(0;x) \propto \frac{1}{u(x)}, & \qquad \qquad  \tilde{E}(0) = \eps.
\end{cases}
\end{align}

\subsection{Ladder operators for the type III rational extensions} \label{sec:3.2_LO_RE}
For most solvable systems, SUSYQM offers a natural way to adapt the ladder operators of the initial system to that of any partner \cite{LO.13-12-23, L0.19-02-25, LO.19-07}. Suppose that ladder operators are known for the initial system. The idea is to use the relation (\ref{eq:1.2_connection}) to perform the raising or lowering action on the eigenstates of the initial system with the known ladder operators, then to make use of the same relation in the opposite direction to return to the SUSY partner system. The composition of these three actions acts as ladder operators on the SUSY partner system. We wish to construct ladder operators $\mathcal{A}^\pm$ for the type III rational extension $\tilde{V}(x)$ of the RMII system according to this technique.

On the other hand, the additional ground state $\tilde{\psi}(0;x)$ is not related to any state of the initial RMII system $H_s$, where the previously constructed $A^\pm$ operates. This fundamental distinction that $\tilde{\psi}(0;x)$ has from the rest of the eigenstates of $\tilde{H}$ in the state-adding SUSY transformation induces a direct sum decomposition of the Hilbert space for the type III rational extensions into two subspaces: one spanned by the states $\tilde{\psi}(n;x)$, $n \geq 1$, in correspondence with $H_s$, and one with the added state $\tilde{\psi}(0;x)$. Such decomposition has been discussed for higher order state-adding SUSY transformations, for instance, in \cite{CS.19-01-11}.

In this sense, the ladder operators obtained from the above mentioned technique are only consistent in the subspace that is in correspondence with the initial system $H_s$, meaning that it  does not allow to construct ladder operators connecting $\tilde{\psi}(0;x)$ to $\tilde{\psi}(1;x)$.  Moreover, the lowering action obtained acting on $\tilde{\psi}(1;x)$ yields the annihilation of the state. Therefore, we use the technique to construct ladder operators $\mathcal{A}^\pm$ acting within the subspace spanned by $\{\tilde{\psi}(1;x), \tilde{\psi}(2;x), \dots, \tilde{\psi}(n_{max};x) \}$ only. In some cases (see, e.g. \cite{CS.19-01-11, LO.19-05-23}), it is possible to construct ladder operators separately in the state-added subspace. But since in our case this subspace is spanned by $\{\tilde{\psi}(0;x)\}$ and contains only one energy level, the idea of a consistent ladder operator is not well-defined.

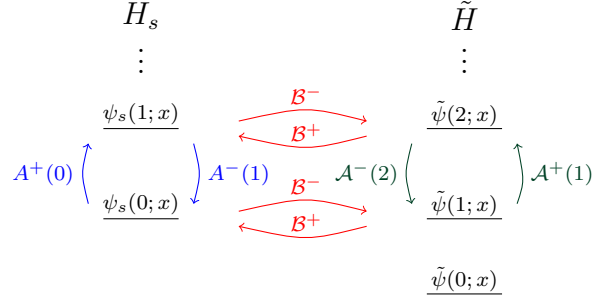
\begin{figure}[h]
\centering
\begin{tikzpicture}

\node at (-0.2,1.0) {\scriptsize $\psi_s(0;x)$};
\node at (-0.2,2.2) {\scriptsize $\psi_s(1;x)$};

\node at (-0.2,3.5) {{$H_s$}};
\node at (-0.2,3) {$\vdots$};
\draw (-0.7,2) -- (0.3,2);
\draw (-0.7,0.8) -- (0.3,0.8);

\node at (4.1,0.0) {\scriptsize $\tilde{\psi}(0;x)$};
\node at (4.1,1.0) {\scriptsize $\tilde{\psi}(1;x)$};
\node at (4.1,2.2) {\scriptsize $\tilde{\psi}(2;x)$};

\node at (4.1,3.5) {{$\tilde{H}$}};
\node at (4.1,3) {$\vdots$};
\draw (3.6,2) -- (4.6,2);
\draw (3.6,0.8) -- (4.6,0.8);
\draw (3.6,-0.2) -- (4.6,-0.2);

\draw[red,->] (1.1,0.9) .. controls (2,1.1) .. (2.8,0.9);
\draw[red,->] (2.8,0.7) .. controls (2,0.5) .. (1.1,0.7);
\node[red] at (2,1.25) {\scriptsize $\mathcal{B}^-$};
\node[red] at (2,0.75) {\scriptsize $\mathcal{B}^+$};
\draw[red,->] (1.1,2.1) .. controls (2,2.3) .. (2.8,2.1);
\draw[red,->] (2.8,1.9) .. controls (2,1.7) .. (1.1,1.9);
\node[red] at (2,2.45) {\scriptsize $\mathcal{B}^-$};
\node[red] at (2,1.95) {\scriptsize $\mathcal{B}^+$};

\draw[blue,->] (-0.9,1.0) .. controls (-1,1.4) .. (-0.9,1.8);
\node[blue] at (-1.5,1.4) {\scriptsize $A^+(0)$};

\draw[blue,->]  (0.5,1.8).. controls (0.6,1.4) ..  (0.5,1.0);
\node[blue] at (1.1,1.4) {\scriptsize $A^-(1)$};

\draw[gg,->] (3.4,1.8) .. controls (3.3,1.4) ..  (3.4,1.0);
\node[gg] at (2.8,1.4) {\scriptsize $\mathcal{A}^-(2)$};

\draw[gg,->] (4.8,1.0) .. controls (4.9,1.3) ..   (4.8,1.8);
\node[gg] at (5.4,1.4) {\scriptsize $\mathcal{A}^+(1)$};
\end{tikzpicture}
\caption{Construction of the ladder operators $\mathcal{A}^\pm$ for the rational extension of the RMII system using $A^\pm$ and $\mathcal{B^\pm}$.}
\label{fig:3.2_ladder_RE}
\end{figure}

Hence, starting from an excited eigenstate $\tilde{\psi}(n;x)$, $n \geq 1$, of the rational extension $\tilde{V}(x)$ of (\ref{eq:3.1_RE}), we first act with $\mathcal{B}^+$ to return to $\psi_s(n-1;x)$ of the initial RMII system. Then, we apply the raising $A^+$ or lowering operators $A^-$ developed in (\ref{eq:2.4_raising}) and (\ref{eq:2.4_GS_lower}) within the initial system to get to $\psi_s(n;x)$ or $\psi_s(n-2;x)$ \footnote{Starting with $\tilde{\psi}(1;x)$, a lowering action will yield the annihilation of the state.}. This switches to the adjacent energy level. We then finally recover the corresponding states $\tilde{\psi}(n\pm1;x)$ of the rational extension after acting with $\mathcal{B}^-$. The composition of these actions produce $\mathcal{A}^\pm$ as illustrated in Figure \ref{fig:3.2_ladder_RE} for the states $\tilde{\psi}(1;x)$ and $\tilde{\psi}(2;x)$.

With this approach, we obtain ladder operators $\mathcal{A}^\pm$ acting on all eigenstates of the type III rational extensions except $\tilde{\psi}(0;x)$:
\begin{align}
\mathcal{A}^\pm(n) &= \frac{\mathcal{B}^- A^\pm(n-1)\ \mathcal{B}^+}{\sqrt{\qty(\tilde{E}(n \pm 1) - \eps) \qty(\tilde{E}(n) - \eps)}}. \label{eq:3.2_lower_RE}
\end{align}
The ladder operator action is realized on that subspace ($n \geq 1$):
\begin{align}
\mathcal{A}^-(n)\ \tilde{\psi}(n;x) = \sqrt{\tilde{k}(n)}\ \tilde{\psi}(n-1;x), \qquad  \mathcal{A}^+(n)\ \tilde{\psi}(n;x) = \sqrt{\tilde{k}(n+1)}\ \tilde{\psi}(n+1;x),
\end{align}
this time with the annihilation $\mathcal{A}^-(1) \tilde{\psi}(1;x) = 0$. The function $\tilde{k}(n)$ is again chosen to be the shifted energy with respect to the lowest excitation considered ($n=1$):
\begin{align}\label{eq:3.2_RE_shifted}
\tilde{k}(n) = \tilde{E}(n) - \tilde{E}(1) = E_s(n-1) - E_s(0) = k(n-1).
\end{align}
The ladder are thus again $2n \pm 1$ order differential operators due to the shift $m = n-1$ occurring in the excitation for a state-adding SUSY transformation, except for $\mathcal{A}^-(1)$, which is of the second order. 

\section{Coherent states for the RMII potential and the type III rational extensions}\label{sec:4}
The Barut-Girardello coherent states are defined as eigenstates of the lowering operator $A^-$ of the system under study. After obtaining ladder operators realizations for the RMII system and for its type III rational extensions, we now use them for the construction of such coherent states. Section \ref{sec:4.1_CS} briefly present the formalism of Barut-Girardello coherent states and how it adapts for finite discrete spectrum systems like the RMII. We further adapt the construction of the coherent states for the type III rational extensions. Then, in Section \ref{sec:4.2}, properties of space-localization, trajectories and position-momentum uncertainty relations are analyzed and compared for the coherent states of the type III rational extensions with respect to the work of \cite{CS.12-10-12} on the RMII system.

\subsection{Coherent states constructions}\label{sec:4.1_CS}
Barut-Girardello coherent states $\phi(w;x)$  are defined as normalizable eigenstates of a lowering operator $A^-$ of a given solvable system \cite{CS.71}. For infinite discrete spectrum systems, the definition 
\begin{align}
A^- \phi(w;x) = w \phi(w;x), \qquad \qquad w \in \mathbb{C},
\end{align}
forces the coherent states to be a superposition of the energy eigenstates $\psi(n;x)$ of the form:
\begin{align}\label{eq:4.1_infinite_sup}
\phi(w;x) = \frac{1}{\sqrt{\mathcal{N}(|w|^2)}} \sum_{n = 0}^\infty \frac{w^n}{\sqrt{\rho(n)}} \psi(n;x),
\end{align}
where $\mathcal{N}(|w|^2)$ ensures normalization and where:
\begin{align}\label{eq:4.1_rho}
\mathcal{N}(|w|^2) =  \sum_{n=0}^{\infty} \frac{|w|^{2n}}{\rho(n)}, \qquad \qquad \rho(n) = \prod_{j=1}^n k(j), \qquad \qquad \rho(0) = 1,
\end{align}
are defined in terms of a function $k(n)$ coming from the ladder operator action (\ref{eq:1.1_LO}). In this sense, the existence and realization of a lowering operator is important in the construction of Barut-Girardello coherent states for the system. 

The superposition (\ref{eq:4.1_infinite_sup}) has been extended in the case of finite spectrum systems \cite{ CS.12-10-12, CS.08-07-15} by terminating the sums in (\ref{eq:4.1_infinite_sup}) and (\ref{eq:4.1_rho}) at $n_{max}$. Due to the finiteness of the superposition in this case, these states are called \textit{almost eigenstates} of the lowering operator:
\begin{align}
A^- \phi(w;x) \approx w \phi(w;x), \qquad \qquad w \in \mathbb{C},
\end{align}
in the sense that the deviation from being an exact eigenstate is small \cite{CS.12-10-12, CS.08-07-15}. According to this definition, the coherent states defined as almost eigenstates of the RMII lowering operator $A^-$ constructed in Section \ref{sec:2} are:
\begin{align}\label{eq:4.1_finite_cs}
\phi_s(w;x) = \frac{1}{\sqrt{\mathcal{N}_s(|w|^2)}} \sum_{n = 0}^{n_{max}} \frac{w^n}{\sqrt{\rho_s(n)}} \psi_s(n;x), \qquad \qquad w \in \mathbb{C},
\end{align}
with $\rho_s(n)$ given through (\ref{eq:4.1_rho}) by $k(n) = E_s(n) - E_s(0)$, the RMII shifted energy appearing in the action of $A^\pm$. In fact, the precise correction needed for $\phi(w;x)$ to be an exact eigenstate has been computed in \cite{CS.12-05-30}. In addition, the RMII potential is used in practice, or serves as a basis, in the study of polyatomic molecules where the allowed number of excited states is $n_{max} \gtrsim 12$ \cite{CS.13-10-29, CS.62-04-01, CS.21-03, CS.08-06-17}, in which case the contribution of the maximally excited eigenstate $\psi_s(n_{max};x)$ in the superposition (\ref{eq:4.1_finite_cs}) is inconsequential compared to that of the other excitations. This superposition of the RMII eigenstates has been studied in \cite{CS.12-10-12}, and their properties will be used to compare with that of the following coherent states construction for the type III rational extensions. 

We adapt this construction for the type III rational extensions of the RMII system to obtain almost eigenstates $\tilde{\phi}(w;x)$ of the lowering operator $\mathcal{A}^-$. Since the lowering operator $\mathcal{A}^-$ does not act on the ground state of the rational extensions, we adapt the superposition (\ref{eq:4.1_finite_cs}) to take eigenstates $\tilde{\psi}(n;x)$ of excitation $n \geq 1$ \cite{CS.07-05-30}. Therefore, the adapted coherent states for the type III rational extensions can be written as:
\begin{align}
\tilde{\phi}(w;x) = \frac{1}{\sqrt{\tilde{\mathcal{N}}(|w|^2)}} \sum_{n=0}^{n_{max}} \frac{w^n}{\sqrt{\tilde{\rho}(n+1)}} \tilde{\psi}(n+1;x),
\end{align}
where:
\begin{align}
\tilde{\mathcal{N}}(|w|^2) =  \sum_{n=0}^{n_{max}} \frac{|w|^{2n}}{\tilde{\rho}(n+1)}, \qquad \qquad \tilde{\rho}(n) = \prod_{j=1}^n \tilde{k}(j), \qquad \qquad \tilde{\rho}(1) = 1,
\end{align}
with $\tilde{k}(n)$ taken from the action of $\mathcal{A}^\pm$ (\ref{eq:3.2_RE_shifted}). Since the energies of the states considered in this superposition are connected through SUSYQM with that of the states considered in the coherent states $\phi_s(w;s)$ of the RMII system, one has the association $\tilde{\rho}(n+1) = \rho_s(n)$ and thus $\tilde{\mathcal{N}}(|w|^2) = \mathcal{N}_s(|w|^2)$.

\subsection{Properties}\label{sec:4.2}
Having Barut-Girardello type coherent states for the RMII potential and for its type III rational extensions, we now analyze some properties. In the following, we consider the case with parameters $s = 20$ and $\lambda = 16$. We focus on the coherent states $\tilde{\phi}(w;x)$ of the type III rational extension obtained from a degree $k = 2$ seed solution and compare it to that $\phi_s(w;x)$ of the initial RMII system for $w \in \mathbb{R}$. We see from (\ref{eq:4.1_finite_cs}) that the almost eigenstate nature of the coherent states can be lost with increasing value of $w$. In this sense, we restrict to a range of $w$ maintaining such nature and we explore the associated properties.

\begin{figure}[h]
    \centering
    \subfigure[$|\phi_s(w;x)|^2$  (RMII)]{\includegraphics[width=0.35\textwidth]{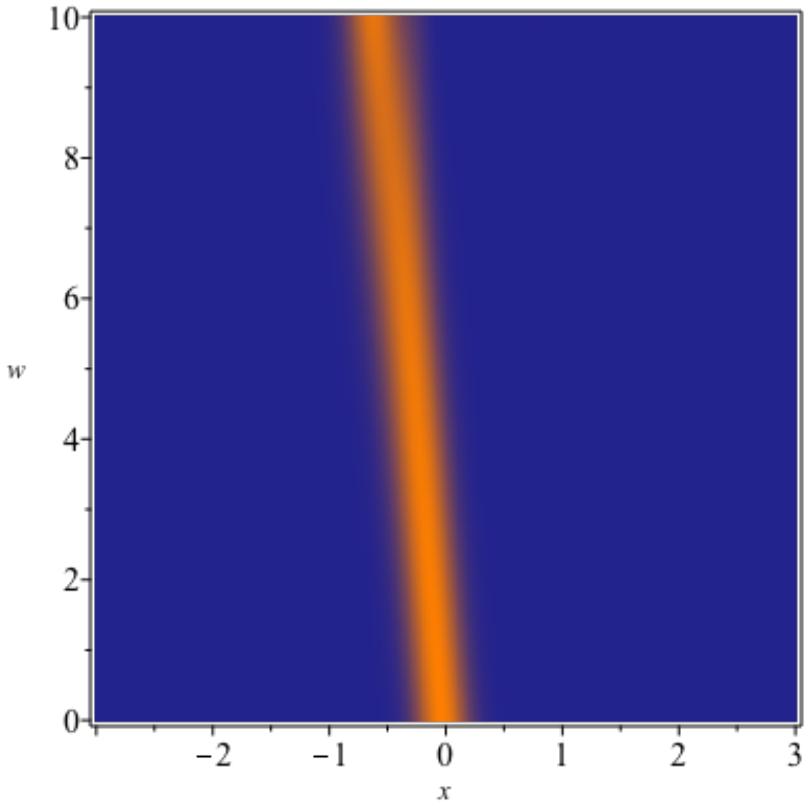}} 
    \hspace{4em}
    \subfigure[$|\tilde{\phi}(w;x)|^2$  (type III)]{\includegraphics[width=0.35\textwidth]{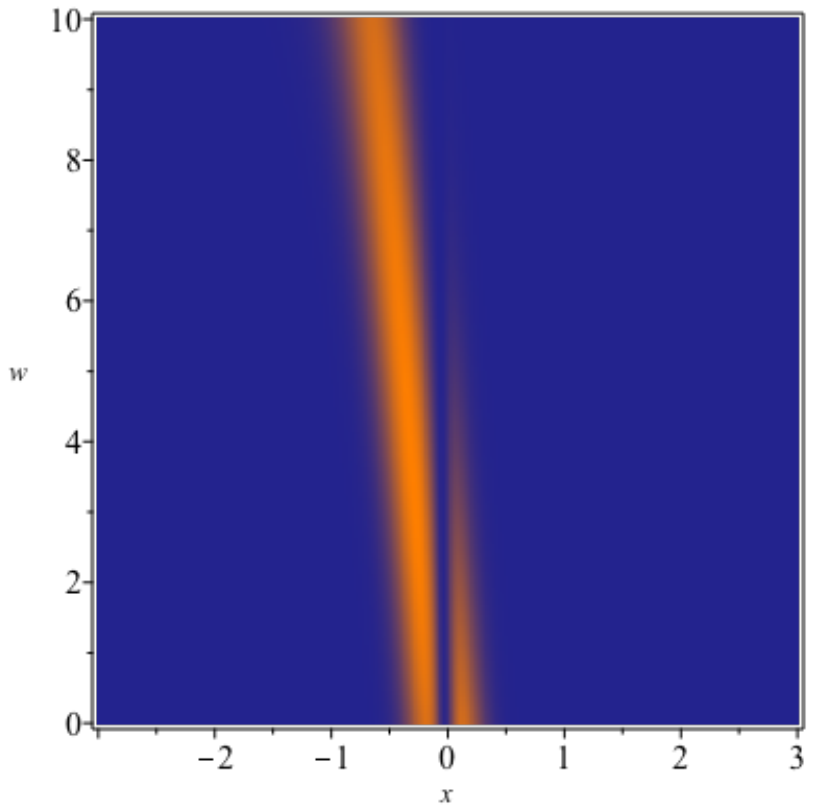}\label{fig:4.2_localization_x-w_RE}}
    \caption{Space localization of the probability densities of the coherent states for $w \in [0,10]$ with parameters $s= 20$ and $\lambda = 16$.}
    \label{fig:4.2_localization_x-w}
\end{figure}

One property that share most of the different types of coherent states is space localization \cite{CS.08-07-15}. Firstly, this is analyzed as a function of the parameter $w \in \mathbb{R}$ for the probability densities $\qty|\phi_s(w;x)|^2$ and $|\tilde{\phi}(w;x)|^2$ exposed in Figure \ref{fig:4.2_localization_x-w}. The probability density of the rational extension allows a secondary maximum that contributes more and more as $w \to 0$. This splitting of the probability density could be explained by the fact that the lowest energy eigenstate considered in the superposition, $\tilde{\psi}(1;x)$, has one node. Since the low energy eigenstates have a larger weight in the superposition for smaller $w$ (especially for $w< 1$), we get a probability density that tends to that of a one-node state when $w \to 0$.

\begin{figure}[h]
    \centering
    \subfigure[$|\Phi(1/2;x,t)|^2$]{\includegraphics[width=0.243\textwidth]{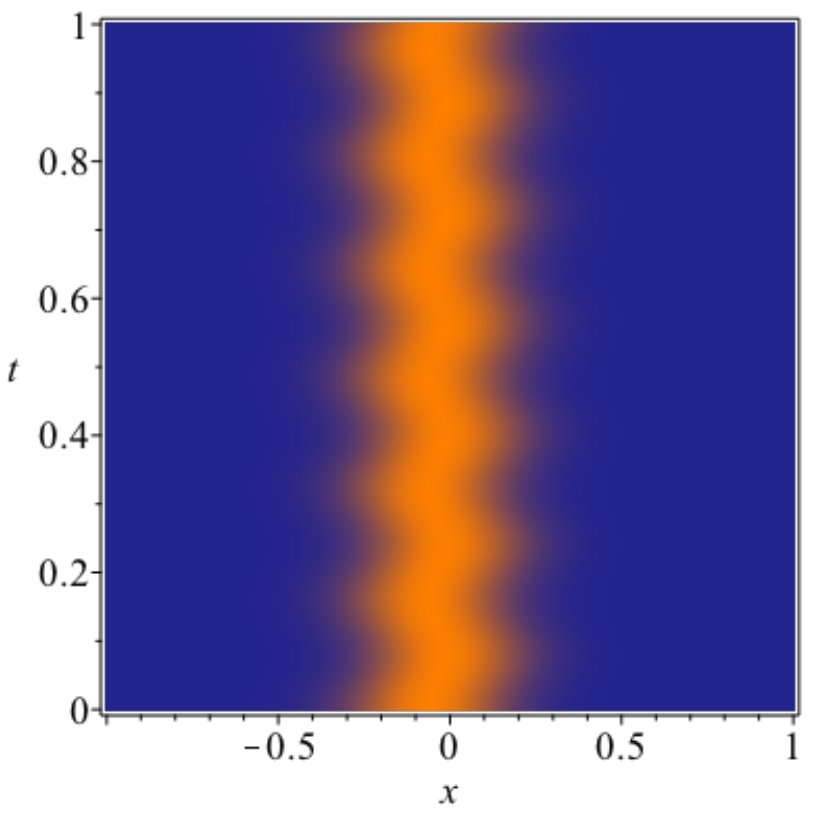}} 
        \subfigure[$|\tilde{\Phi}(1/2;x,t)|^2$]{\includegraphics[width=0.243\textwidth]{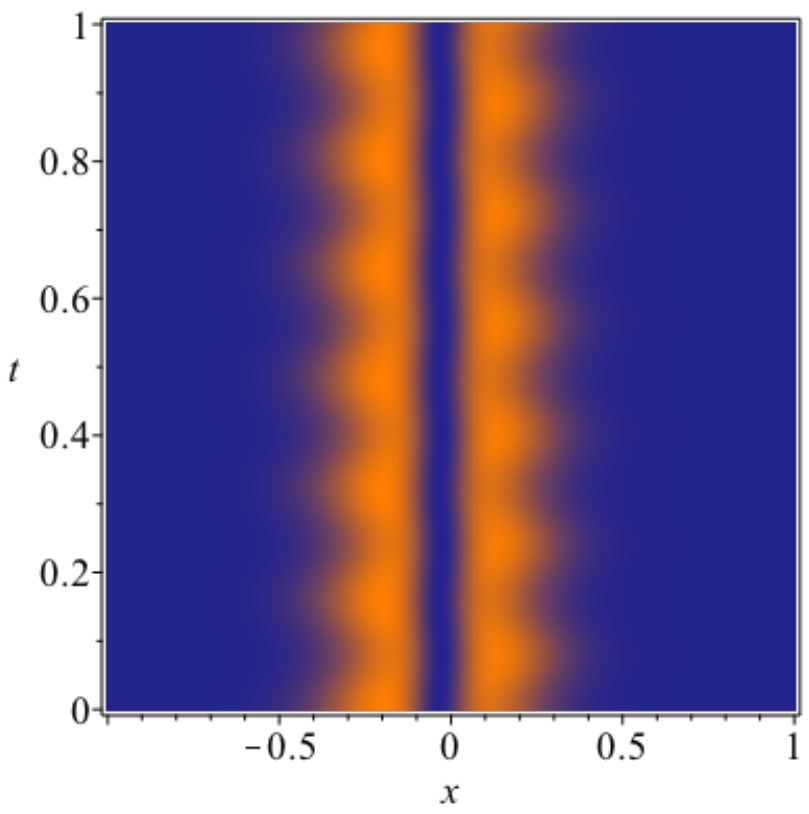}} 
    \subfigure[$|\Phi(5;x,t)|^2$]{\includegraphics[width=0.243\textwidth]{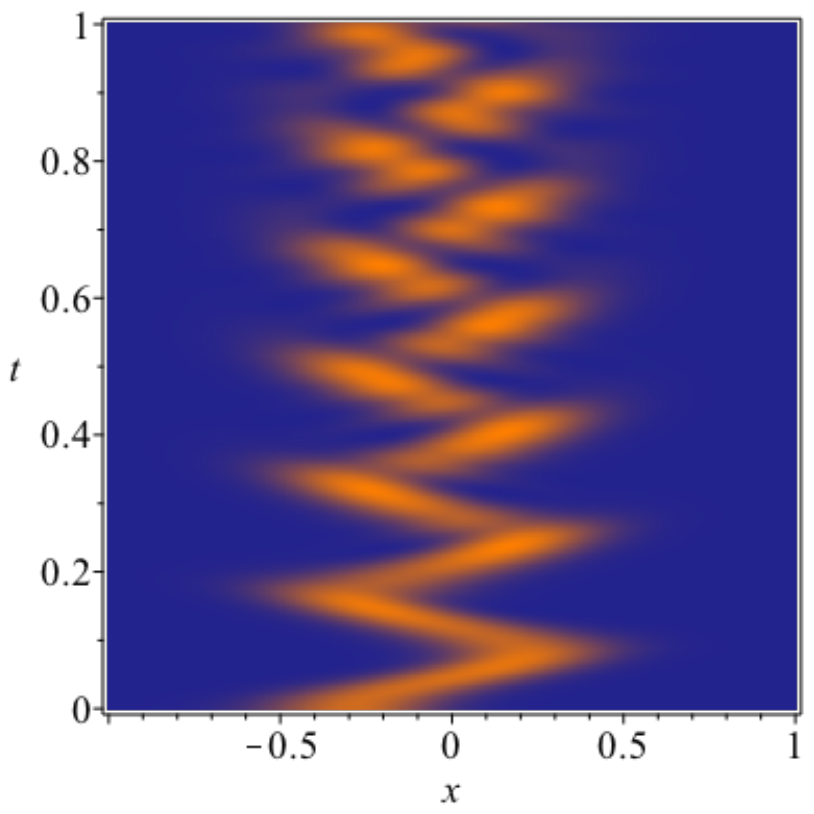}}
    \subfigure[$|\tilde{\Phi}(5;x,t)|^2$]{\includegraphics[width=0.243\textwidth]{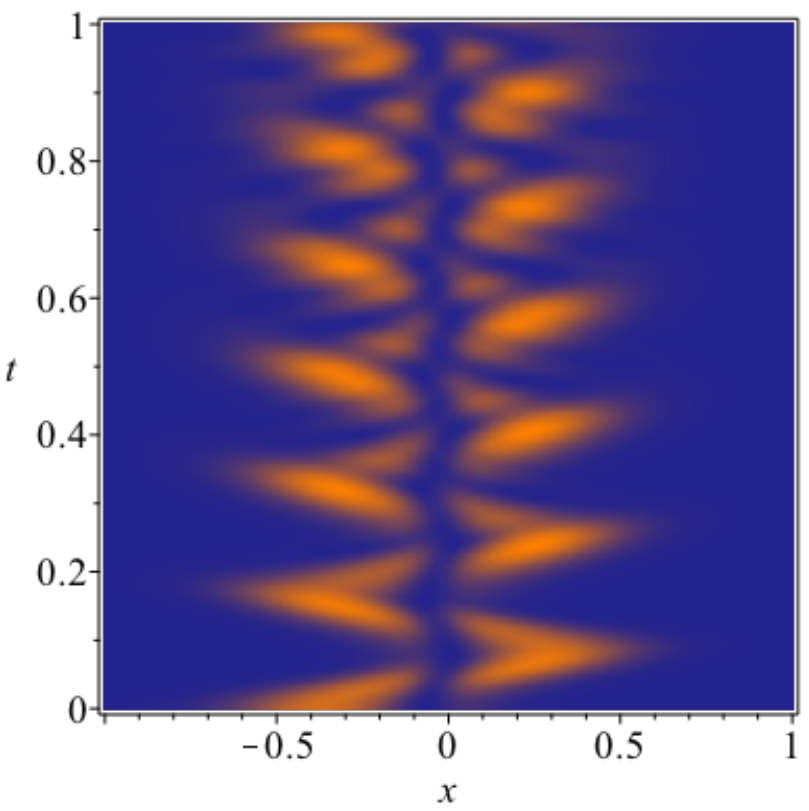}}
    \caption{Space localization of the probability densities $|\Phi(w;x,t)|^2$ and $|\tilde{\Phi}(w;x,t)|^2$of the coherent states in time $t \in [0,1]$ for fixed $w$ with parameters $s = 20$ and $\lambda = 16$.}
    \label{fig:4.2_localization_xt}
\end{figure}

We then analyze time evolution of the space localization of the probability densities for $w$ fixed. Indeed, the time evolved coherent states denoted $\Phi_s(w;x,t)$ and $\tilde{\Phi}(w;x,t)$ are obtained in the standard way by acting with the unitary time evolution operators $\exp(-i H_s t)$ and $\text{exp}(-i \tilde{H} t)$ respectively on $\phi_s(w;x)$ and $\tilde{\phi}(w;x)$ \cite{Dirac.P.A.M}. Time evolution of the probability densities for the two coherent states are presented in Figure \ref{fig:4.2_localization_xt} for $w = 1/2$ (a)-(b) and $w = 5$ (c)-(d). Except for the node in the probability density $|\tilde{\Phi}(w;x,t)|^2$, behaviours are similar in time for both constructions. Frequency are similar since the states of same energy in the SUSY correspondence have the same weight in both constructions. We remark a smaller oscillation amplitude for $w = 1/2$ compared to $w = 5$. Moreover, it is seen that the space localization is lost as $t \to 1$ for both coherent states when $w = 5$, while it seems less affected for $w = 1/2$ over that same period of time.

\begin{figure}[h]
    \centering
    \subfigure[RMII $w = \frac{1}{2}, 1, 2$]{\includegraphics[width=0.2405\textwidth]{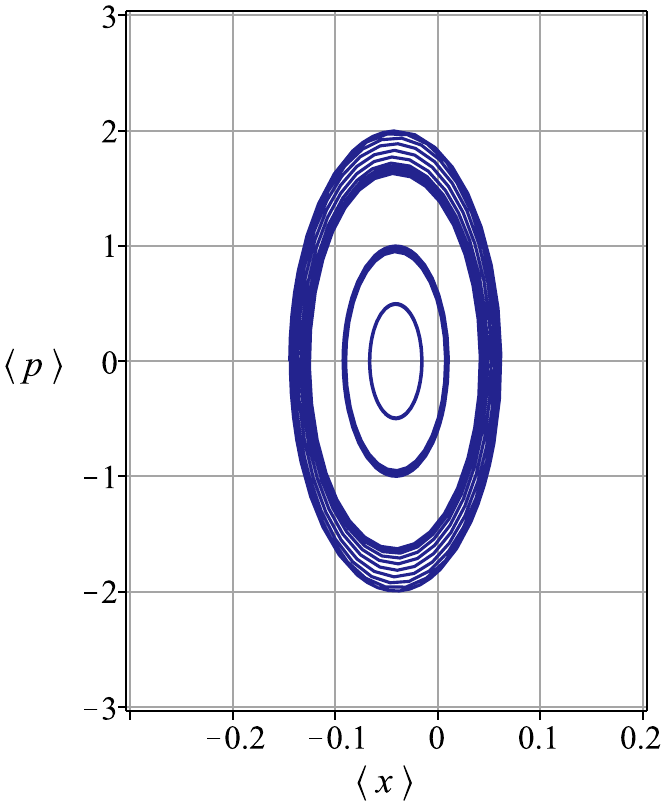}\label{fig:4.2_traj_CS_RM}} 
    \subfigure[Classical RMII]{\includegraphics[width=0.248\textwidth]{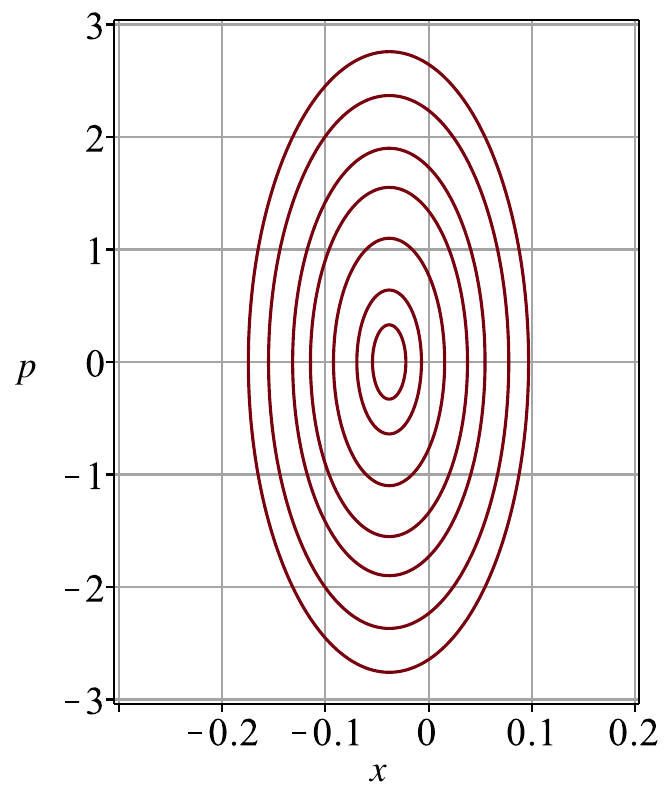}\label{fig:4.2_traj_Class_RM}}
    \subfigure[Type III $w = \frac{1}{2}, 1,2$]{\includegraphics[width=0.2405\textwidth]{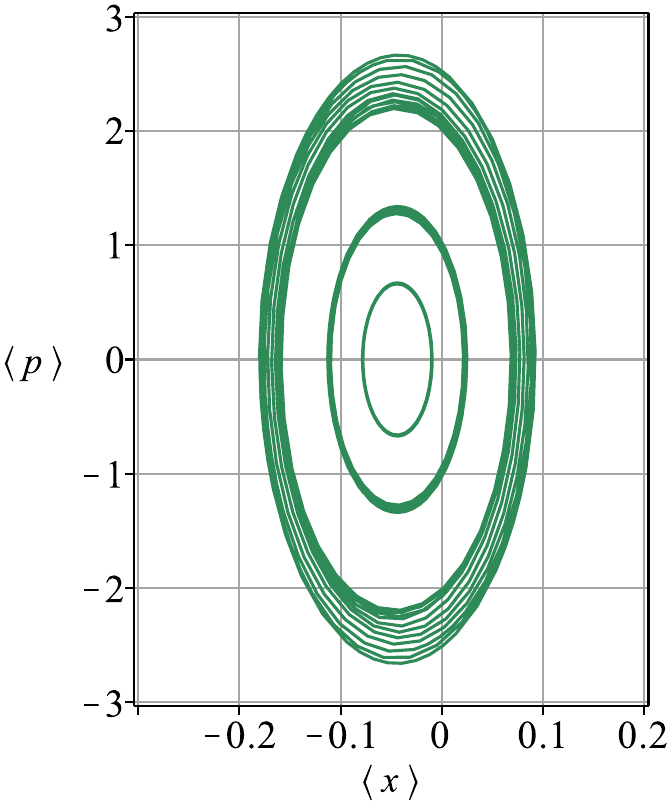}\label{fig:4.2_traj_CS_RE}} 
    \subfigure[{Classical type III}]{\includegraphics[width=0.2405\textwidth]{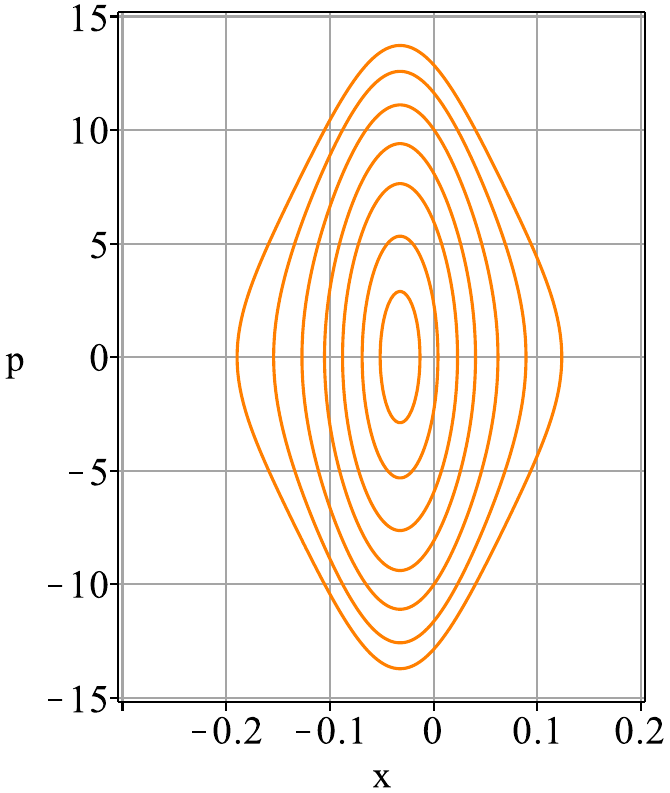}\label{fig:4.2_traj_Class_RE} }
    \caption{Trajectories $\qty( \expval{x}(w;t) , \expval{p}(w;t))$ of position and momentum expectation values in time $t \in [0,3]$ for the coherent states (a)-(c) for $w$ fixed compared to phase space trajectories $(x(t), p(t))$ of bounded motion of their classical analogues (b)-(d) . Parameters: $s=20$ and $\lambda = 16$. }
    \label{fig:4.2_trajectories}
\end{figure}

It is known that for the construction (\ref{eq:4.1_finite_cs}), the trajectory $(\expval{x}(w;t), \expval{p}(w;t))$ of the expectation values of position $x$ and momentum $p$ tends to mimics the phase space trajectory $(x,p)$ describing bounded motion of the classical analogue of the quantum system. This has been verified in \cite{CS.12-10-12} for the RMII coherent states. We investigate how the trajectory of the coherent states of the rational extension compares with respect to that of RMII. Figure \ref{fig:4.2_traj_CS_RM} and \ref{fig:4.2_traj_CS_RE} expose the trajectory for the coherent states of the RMII and of the type III rational extension respectively, for different values of $w$. We have $w = 2$ (exterior), $w = 1$ (middle) and $w = 1/2$ (interior) for the three trajectories in each plot. When comparing to Figure \ref{fig:4.2_traj_Class_RM}, we see that both coherent states have classical RMII-like trajectories. We also see from Figure \ref{fig:4.2_traj_Class_RE} that the trajectory the coherent states of the rational extension does not approximate that of its own classical analogue which has greater amplitude in momentum due to the deep and narrow sub-well of $\tilde{V}(x)$. Indeed, by neglecting the ground state in the construction, the trajectory obtained is expected to be that of a classical bounded motion in a potential $\tilde{V}(x)$ for which that sub-well is removed. Such potential would be slightly deeper than the initial RMII potential $V_s(x)$ (see Figure \ref{fig:3.1_graph_RE}), explaining the greater range in momentum that the trajectory of the rational extension coherent states have with respect to that of the RMII.

\begin{figure}[h]
    \centering
\includegraphics[width=0.50\textwidth]{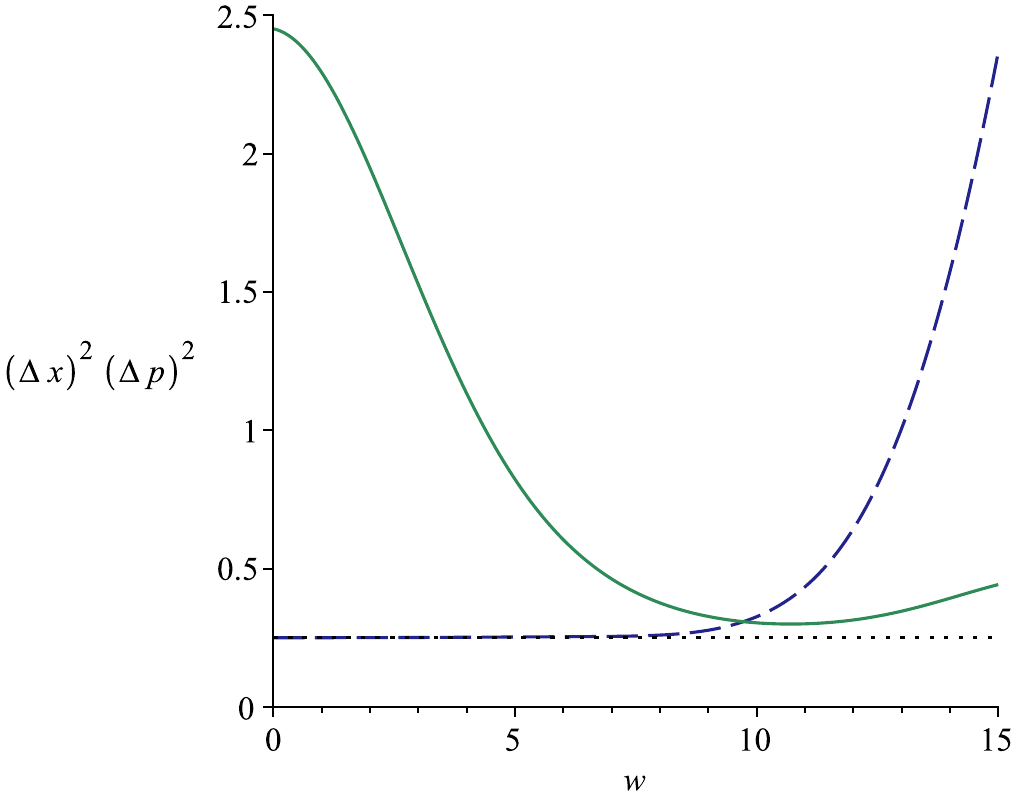}
    \caption{Position-momentum uncertainty relation $(\Delta x)^2 (\Delta p)^2$ at $t = 0$ for the RMII coherent states (blue, dashed) and for that of the type III rational extension (green, solid) as a function of $w \in [0,15]$. Parameters: $s = 20$ and $\lambda = 16$. A black dotted line indicates the lower bound of $1/4$.}\label{fig:4.2_UR}
\end{figure}

The last property we analyze is the minimization of Heisenberg's uncertainty relation for position and momentum:
\begin{align}
(\Delta x)^2(\Delta p)^2 \geq \frac{1}{4}, \qquad \qquad \qquad \Delta \mathcal{O} = \sqrt{\expval{\mathcal{O}^2} - \expval{\mathcal{O}}^2},
\end{align}
which is realized for certain types of coherent states, but not necessarily for Barut-Girardello types \cite{CS.12-10-12}. We numerically computed the uncertainty $(\Delta x)^2 (\Delta p)^2 $ at $t=0$ as a function of the parameter $w$ for the coherent states of the RMII system and of its type III rational extension and displayed them in Figure \ref{fig:4.2_UR}. For the small $w$ studied regime, the RMII coherent states approximate the lower bound of the uncertainty relation, but begin to diverge from it at around $w =9$, where the almost eigenstate status of the coherent state is less and less realized. For the coherent states of the type III rational extension, the non-approximation of the lower bound for small $w$ is in correspondence with the amplitude of the second maxima in the probability density $|\tilde{\phi}(w;x)|^2$ shown in Figure \ref{fig:4.2_localization_x-w_RE}. Once this second maximum is negligible (about $w \in [9,13]$), the uncertainty relation is better minimized, before diverging away from the lower bound at larger $w$.

\section{Generalization to the RMI potential}\label{sec:5}

This section intends to generalize the ladder operators construction and coherent states to the trigonometric Rosen-Morse (RMI) potential \cite{ tRM.05-12-21, tRM.08-10-20, tRM.10-10-14}. It is known that both RMI and RMII systems can each be mapped onto the other by a point canonical transformation (PCT) \cite{SS.92-02-13, tRM.62-04-24, tRM.20-02-28}. We summarize the transition from the RMII to the RMI system in Section \ref{sec:5.1}. We will label the different systems by $I$ and $II$ accordingly. We show how the ingredients for the ladder operators construction transform and obtain analogue ladder operators for this system in Section \ref{sec:5.2}. Once all necessary components of the ladder operators construction have been recovered in the RMI setting, we will drop the $I$ and $II$ labeling and transfer our usual $s$ labeling to the RMI system for the rest. The associated Barut-Girardello coherent states are then analyzed in Section \ref{sec:5.3}.

\subsection{The RMI system from a point canonical transformation}\label{sec:5.1}
\begin{figure}[h]
\centering
\includegraphics[width=0.40\textwidth]{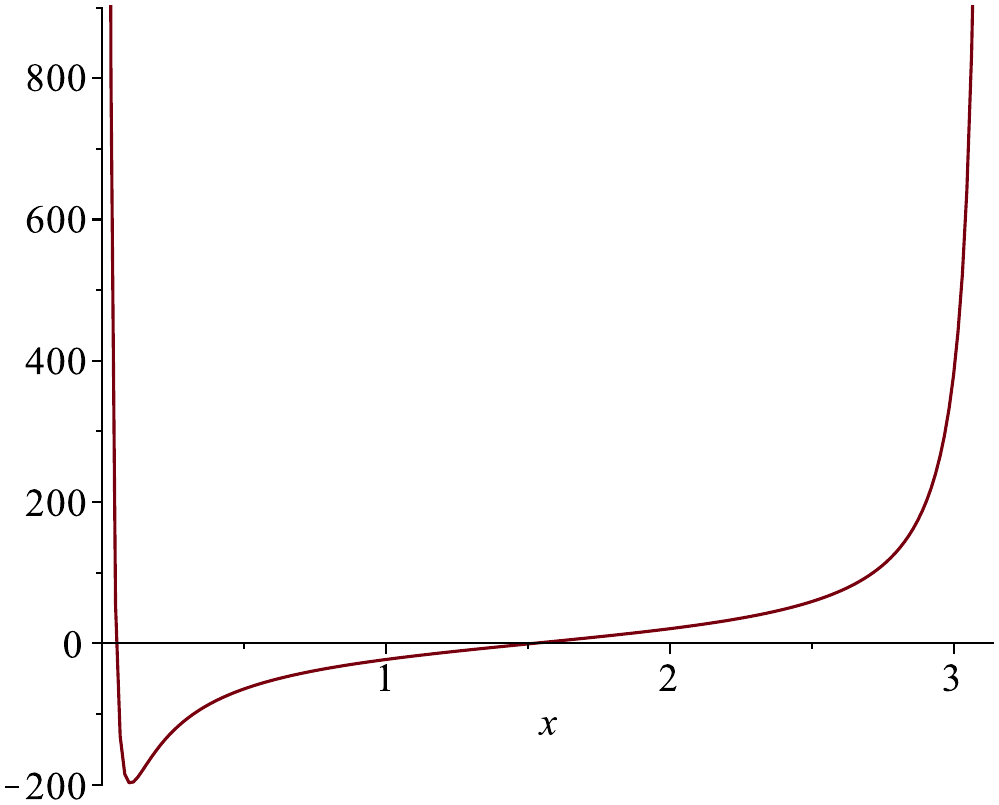}
\caption{Trigonometric Rosen-Morse (RMI) potential $V_I(x)$ with parameters $s = 2$ and $\lambda = 20$.}\label{fig:6.1_RMI}
\end{figure}

In quantum mechanics, PCT is a method that allows to transform a given Schrödinger equation into a new one when a change of variables is applied \cite{SS.92-02-13, tRM.62-04-24, tRM.20-02-28}. We summarize how this is done for the Rosen-Morse potentials. We label by $V_{II}(x)$ the hyperbolic Rosen-Morse (RMII) potential defined in Section \ref{sec:2.1_RMII} and label accordingly by $\psi_{II}(n;x)$ and $E_{II}(n)$ its bounded eigenstates and energies. The Schrödinger equation for the normalizable eigenstates of this system can be written as
\begin{align}\label{eq:5.1_PCT_SE_RMII}
\qty(-\dv[2]{x} + V_{II}(x) - E_{II}(n)) \psi_{II}(n;x) = 0.
\end{align}
The PCT is performed through the following complex transformation of variable and parameters \cite{tRM.20-02-28, tRM.13-12-03}:
\begin{align}\label{eq:5.1_PCT}
x \to ix + \frac{i \pi}{2}, \qquad \qquad \lambda \to i \lambda, \qquad \qquad  s \to -s.
\end{align}
Equation (\ref{eq:5.1_PCT_SE_RMII}) becomes:
\begin{align}
\qty(\dv[2]{x} + 2 \lambda \cot{x} - s(s-1)\csc^2{x} + (s-n)^2 - \frac{\lambda^2}{(s+n)^2})   \psi_{II}\qty(n; ix+  \frac{i \pi}{2}) = 0,
\end{align}
where the parameters $\lambda$ and $s$ are changed in the eigenstate according to (\ref{eq:5.1_PCT}).  In order to associate the later to a Schrödinger equation, we have to multiply the operator factor by $-1$. We furthermore restrict the domain of the new spatial variable to $x \in [0, \pi]$ due to divergences and periodicity. This way, the Schrödinger equation for the trigonometric Rosen-Morse (RMI) potential $V_I(x)$ is recovered:
\begin{align}
\qty(-\dv[2]{x} + V_I(x)- E_{I}(n)) \psi_I(n;x) = 0, \qquad \qquad x \in [0, \pi],
\end{align}
where the potential takes the form \cite{Cooper.F, tRM.05-12-21, tRM.08-10-20, tRM.10-10-14}:
\begin{align}
V_{I}(x) = -2 \lambda \cot{x} + s(s-1) \csc^2{x}.
\end{align}
We demand $s > 1$ to ensure $V_I(x)$ to be a well, and similarly restrict to $\lambda \geq 0$ without loss of generality. Figure \ref{fig:6.1_RMI} illustrates the case for $s = 2$ and $\lambda = 20$. 

The transformed eigenstates and energies take the form:
\begin{align}\label{eq:5.1_RMI_states}
\psi_I(n;x) = K(n) \sin^{s+n}(x)\ \e^{-\frac{\lambda x}{s+n}}\ i^n P^{(c(n),d(n))}_n(i \cot{x}), \quad  E_I(n) = (s+n)^2 - \frac{\lambda^2}{(s+n)^2},
\end{align}
where the analogue parameters $c(n)$ and $d(n)$ and the normalization constant $K(n)$ are now \cite{tRM.08-10-20}:
\begin{align}
c(n) = -(s+n) + \frac{i \lambda}{s+n}, \qquad \ \qquad d(n) = -(s+n) - \frac{i \lambda}{s+n}, \\
K(n) = \e^{\frac{\lambda \pi}{2(s+n)}} 2^{n+s} \sqrt{\frac{n!  \qty((s+n)^2 + \frac{\lambda^2}{(s+n)^2})  \qty| \Gamma \qty(s +\frac{i \lambda}{s+n}) |^2}{\pi  (2s + 2n) \Gamma(2s+n)}} .
\end{align}
Through the PCT, the modification of the potential results here in a infinite bounded spectrum ($n \in \mathbb{Z}^{\geq 0}$), as opposed to the hyperbolic case. Despite the presence of complex parameters and arguments, all eigenstates $\psi_I(n;x)$ are purely real. An expression involving only real parameters and arguments can be given in terms of the Romanovski polynomials $R^{(\alpha,\beta)}_n(y)$ and is developped in \cite{tRM.05-12-21, tRM.13-12-03, tRM.07-09}, for instance.

\subsection{Ladder operators for the RMI potential}\label{sec:5.2}
In this section we construct the ladder operators for the RMI system by recovering every component of the ladder operators construction from the point canonical transformation presented in the previous section.

We first investigate how to recover a hierarchy of RMI systems. We use the Hamiltonian factorization realized by the intertwining operators of SUSYQM (\ref{eq:1.2_facto}). Using $II$ or $I$ as indices to distinguish between systems, we indeed start with the result from the state-deleting SUSY:
\begin{align}
H_{II} - E_{II}(0) = B^+_{II} B^-_{II}.
\end{align}
The LHS is precisely the operator factor of (\ref{eq:5.1_PCT_SE_RMII}). To get to the trigonometric Hamiltonian $H_I$, we apply the transformation (\ref{eq:5.1_PCT}) and multiply by $-1 = (-i) \cdot (-i)$. Therefore, the intertwining operators of the RHS transform according to :
\begin{align}
i B^\pm_{II}
\xrightarrow[ \scriptsize x \to ix + i \pi/2 ]{ \scriptsize \mqty{ \lambda \to i \lambda \\ s \to -s}} 
B^\pm_{I},
\end{align}
and take precisely the form:
\begin{align}
B^\pm_{I} = s \cot{x} - \frac{\lambda}{s} \pm \dv{x}.
\end{align}
The shape invariance property of $V_{I}(x)$ is inherited from that of $V_{II}(x)$, but the passage $s \to -s$ in the PCT results in the translation of the parameter to now become $s \to s+1$ at each state-deleting SUSY transformation creating the hierarchy \cite{tRM.10-10-14, tRM.89-03-21}. Now that the transition to the trigonometric system is completed and that shape invariance properties are established, we drop the index $I$ from now on and use the notation of Section \ref{sec:2} (labeling with $s$) without ambiguity on the trigonometric Rosen-Morse system.

The invertible ground state connection is inferred from the transformation of the ground states of each member of the hierarchy:
\begin{align}
\psi_{s+1}(0;x) = \qty(\frac{K_{s+1}(0)}{K_s(0)} \gamma_s(x))\psi_s(0;x), \qquad \qquad \gamma_{s}(x) = \sin{x}\  \e^{-\frac{\lambda x}{s(s+1)}}.
\end{align}

Then, the ladder operators for the trigonometric RM system are constructed in the same fashion as done in Section \ref{sec:2.3_Construction_LO}:
\begin{align}
A^+(n) = B^+_s \qty(\prod_{i=1}^n \frac{B^+_{s+i}}{\sqrt{E_s(n+1) - E_s(i)}}) \qty(\frac{K_{s+n+1}(0)}{K_{s+n}(0)}\gamma_{s+n}(x)) \qty(\prod_{i=1}^n \frac{B^-_{s+n-i}}{\sqrt{E_s(n)-E_s(n-i)}}), \label{eq:5.1_RMI_lowering}\\
A^-(n) = \qty(\prod_{i=0}^{n-2} \frac{B^+_{s+i}}{\sqrt{E_s(n-1) - E_s(i)}}) \qty(\frac{K_{s+n-1}(0)}{K_{s+n}(0)}\gamma^{-1}_{s+n-1}(x)) \qty(\prod_{i=1}^{n-1} \frac{B^-_{s+n-i}}{\sqrt{E_s(n)-E_s(n-i)}}) B^-_s. \label{eq:5.1_RMI_raising}
\end{align}
Hence, the ladder operators $A^\pm(n)$ are differential operators of order $2n \pm 1$ except for $A^-(0)$ which is also of the first order. The ladder operators action (\ref{eq:1.1_LO}) is realized with $k(n)$ being now the RMI shifted energy.

\subsection{Coherent states of the RMI potential}\label{sec:5.3}

As presented in Section \ref{sec:4.1_CS}, the ladder operators (\ref{eq:5.1_RMI_lowering})-(\ref{eq:5.1_RMI_raising}) for the RMI potential motivate a coherent states construction. In this case, we recall that the trigonometric Rosen-Morse system has an infinite discrete spectrum. Hence, we construct new exact Barut-Girardello coherent states $\phi(w;x)$ from (\ref{eq:4.1_infinite_sup})-(\ref{eq:4.1_rho}) for $k(n) = E_s(n) - E_s(0)$ accordingly being the RMI shifted energy (\ref{eq:5.1_RMI_states}). Other coherent states constructions for this system have been investigated, for instance, in \cite{SY.12-05-30, CS.08-02-15, CS.11-01-11}. 

\begin{figure}[h]
    \centering
    \subfigure[{$|\phi(w;x)|^2$ for $w \in [0,5]$}]{\includegraphics[width=0.3\textwidth]{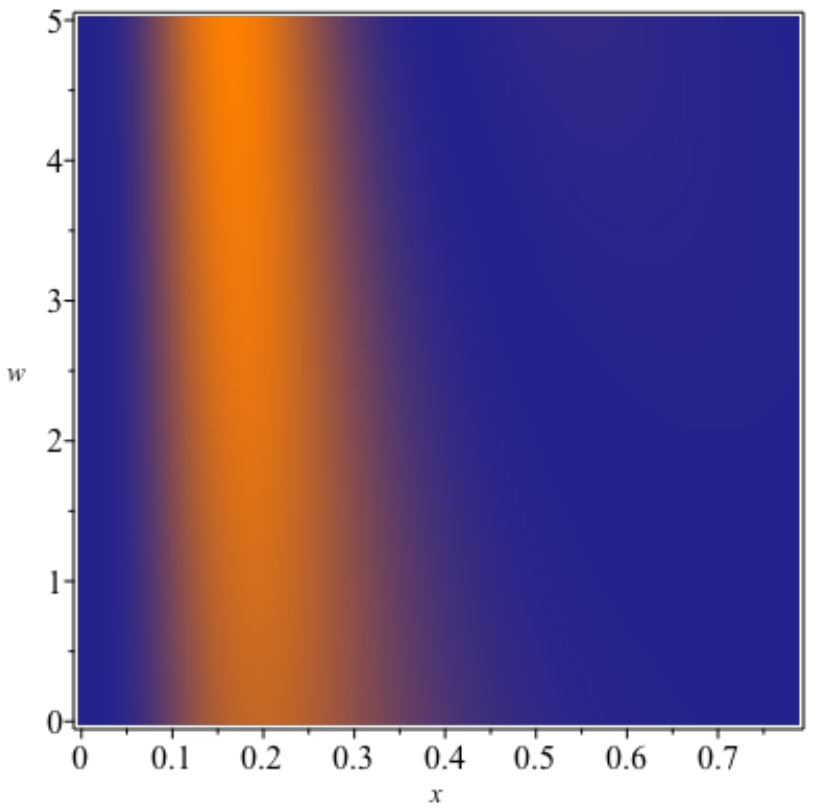}\label{fig:5.3_loc_xw}} 
    \hspace{1em}
    \subfigure[{$|\Phi(1/2;x,t)|^2$ for $t \in [0,1]$}]{\includegraphics[width=0.3\textwidth]{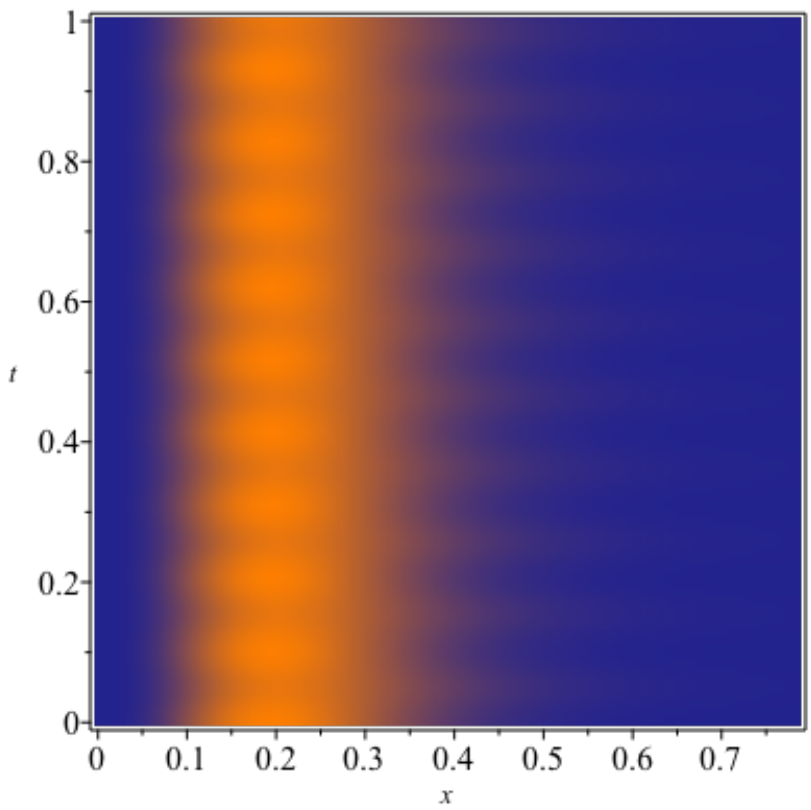} \label{fig:5.3_loc_xt_1_2}}
    \hspace{1em}
    \subfigure[{$|\Phi(2;x,t)|^2$ for $t \in [0,1]$}]{\includegraphics[width=0.3\textwidth]{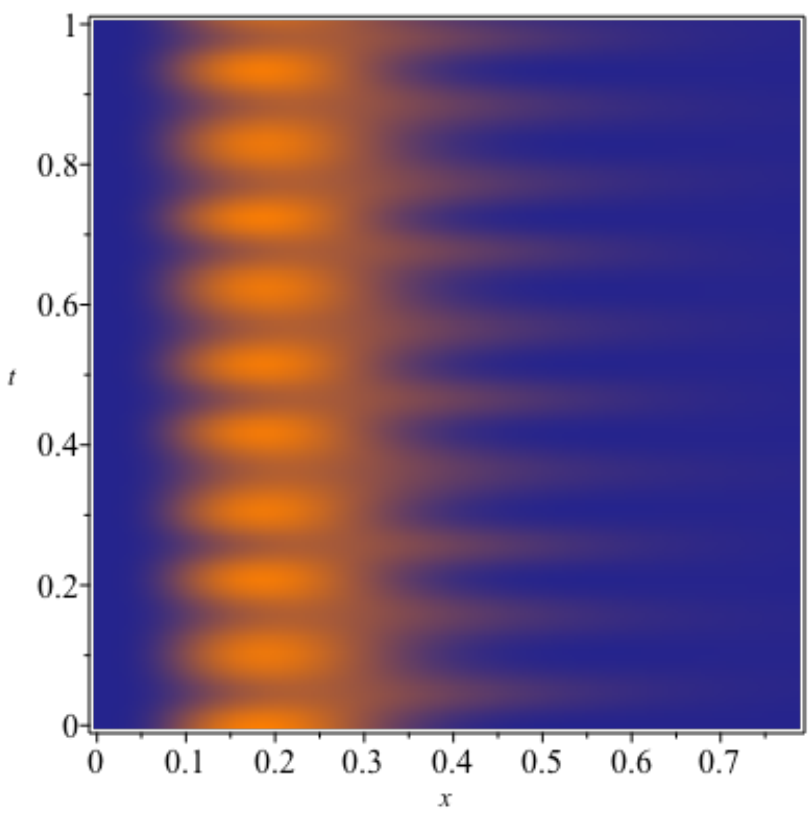} \label{fig:5.3_loc_xt_2}}
    \caption{Space localization of the probability density $|\phi(w;x)|^2$ of the coherent states at $t=0$ in terms of the real parameter $w$ (a). Time evolution of the probability densities $|\Phi(w;x,t)|^2$ for $w$ fixed (b)-(c). Parameters: $s = 2$ and $\lambda = 20$.}
    \label{fig:5.3_localization}
\end{figure}

Space localization of the probability density is exposed in Figure \ref{fig:5.3_localization} for the coherent states of a RMI potential of parameters $s = 2$ and $\lambda = 20$. The localization of $|\phi(w;x)|^2$ at $t=0$ appears in Figure \ref{fig:5.3_loc_xw} from which we remark a low-amplitude secondary maximum appearing around $x \approx 0.6$ as $w$ increases. This secondary maximum is negligible for $w<1$, where the coherent states are most localized. We further analyze the time evolution of the probability densities $|\Phi(w;x,t)|^2$ for $w = 1/2$ in Figure \ref{fig:5.3_loc_xt_1_2} and $w = 2$ in \ref{fig:5.3_loc_xt_2}. Space localization is best preserved in time for $w < 1$. For both values of $w$, the oscillating behaviour of the probability density is uniform for $t \in [0,1]$ as opposed to that of the RMII in Figure \ref{fig:4.2_localization_xt}.

\begin{figure}[h]
    \centering
     \subfigure[RMI  $w=\frac{1}{2},1,2$]{\includegraphics[width=0.35\textwidth]{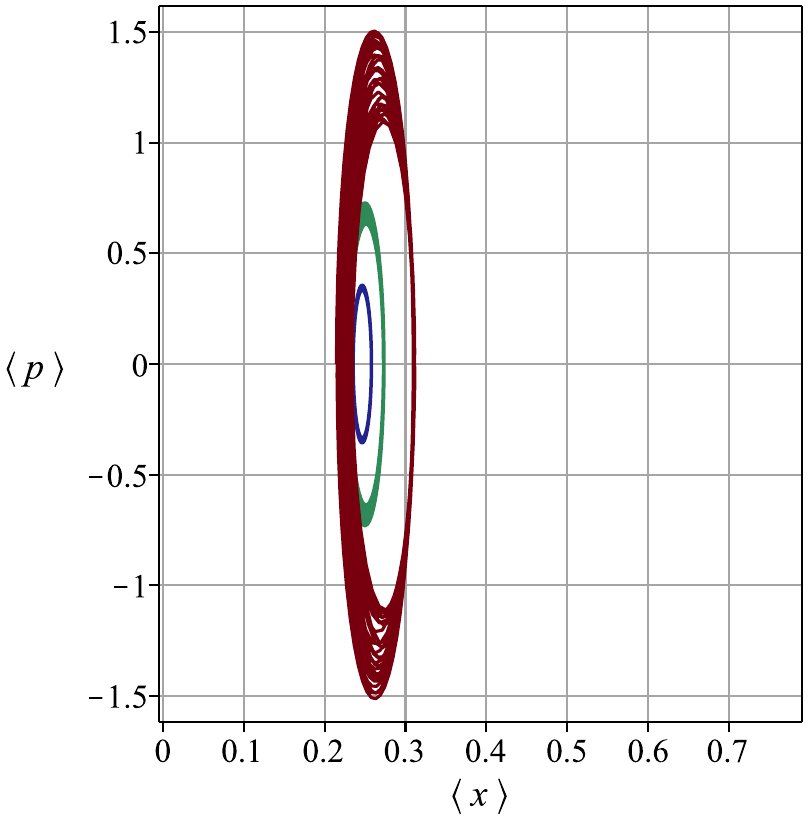}} 
     \hspace{4em}
       \subfigure[RMI classical bounded motion]{\includegraphics[width=0.35\textwidth]{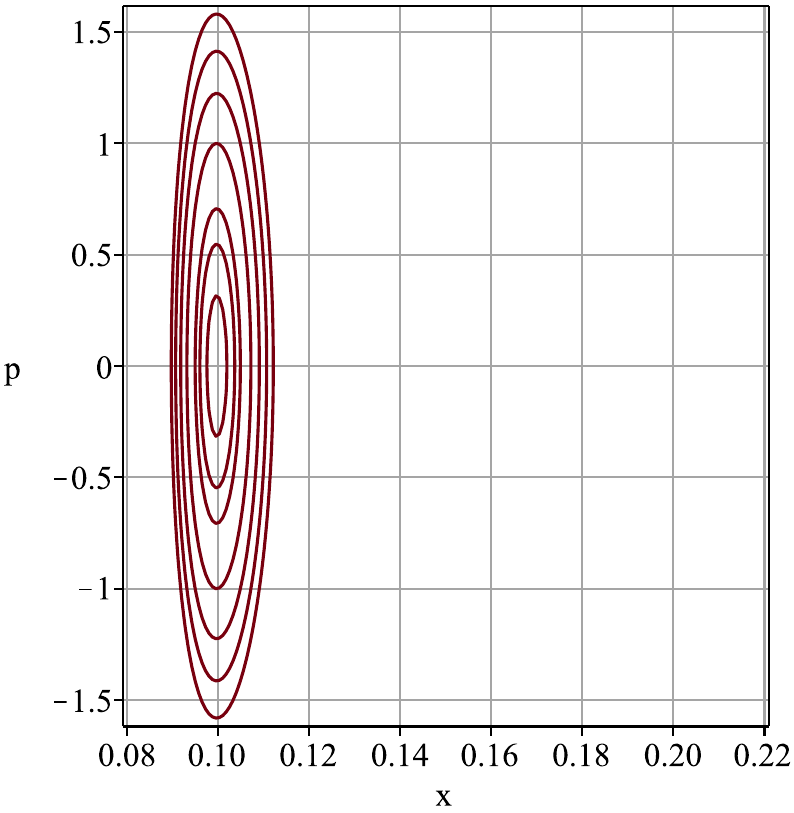}}   
    \caption{Trajectories $\qty( \expval{x}(w;t) , \expval{p}(w;t))$ of position and momentum expectation values in time $t \in [0,3]$ for the coherent states for $w$ fixed (a) compared to phase space trajectories $(x(t), p(t))$ of bounded motion of its classical analogue (b). Parameters: $s=2$ and $\lambda = 20$.}
\label{fig:5.3_trajectory}
\end{figure}

The trajectory $(\expval{x}(w;t), \expval{p}(w;t))$ is computed numerically for different values of $w$ and compared to the phase space trajectory of bounded motion of the classical system in Figure \ref{fig:5.3_trajectory}. We have $w = 2$ (exterior), $w=1$ (middle) and $w = 1/2$ (interior). The trajectories do not agree well, as the classical trajectory is compressed spatially and translated towards to deeper part of the well. An explanation for this could be that only a finite number of eigenstates have their energies in the deep left part of the well, and are thus constrained in the deeper left part of $V_s(x)$. The rest of the eigenstates in the infinite superposition have their probability density spread over the whole domain and would overcome the previously mentioned ones in a way to obtain an effective trajectory whose mean position is farther right with respect to the classical one.

\section{Conclusion and outlook}\label{sec:7}
In this paper, we have constructed ladder operators for the hyperbolic Rosen-Morse (RMII) quantum system whose energies are given by a rational function of the excitation number $n$. Unlike usual solvable systems for which ladder operators are realized as first order differential operators using traditional methods, we used the shape invariance property in SUSYQM together with a ground state connection to recover $A^\pm$ for the RMII potential as $2n \pm1$ order differential operators. These ladder operators were derived completely from a quantum mechanical standpoint, and thus stand apart from the classical approach proposed in \cite{LO.20-06-18}. The choice of $k(n)$ being the shifted energy was natural from construction since the shifted energy appears as proportionality factor in SUSYQM. As an outlook, one could investigate the reducibility of the ladder operators to lower orders, in particular by iteratively making use of the Schrödinger equation to substitute second order derivatives as they act on an eigenstate. Ideas of \cite{LO.09-12} could be useful in that direction.

We have shown that the construction is well suited to be transferred to SUSY partners of the RMII system by constructing that $\mathcal{A}^\pm$ of its type III rational extensions using standard methods. Due to the additional energy level, we restricted the ladder operations on the excited subspace ($n \geq 1$) that is in SUSY correspondence to the inital RMII system. The technique generalizes to other SUSY partners and ladder operators could be constructed for higher-order SUSY partners \cite{SY.20-02-25}, for instance.

In both cases, we used the realization of the ladder operators to construct coherent states as almost eigenstates of the lowering operator. We compared their behaviours with respect to space localization, trajectory and minimization of the Heisenberg uncertainty principle. In a similar fashion to what was done in \cite{CS.12-10-12}, our constructions could be generalized to squeezed states. In this case, the issue of acting with the raising operator on the maximally excited state would have to be addressed, for example, by removing that state from the superposition.

Finally, point canonical transformation has proven to be a successful method to extend our constructions and results to the trigonometric Rosen-Morse (RMI) system. Ladder operators and associated exact Barut-Girardello coherent states we constructed similarly. An outlook would be to fully exploit the PCT network for shape invariant potentials \cite{SS.92-02-13, tRM.20-02-28} to seek ladder operators for other solvable systems. The remaining Kepler-Coulomb potentials are such examples \cite{LO.19-06}. These connections could offer a new approach to tackle the reducibility problem for the ladder operators. Moreover, the classification of the RMI rational extensions have been done in \cite{tRM.13-12-03} using Romanovski polynomials, offering the possibility for an analogous treatment to what was done in this work, this time for type III RMI rational extensions.

\section*{Acknowledgements}
V. Hussin acknowledges the support of research grant from NSERC of Canada. S. Garneau-Desroches acknowledges the Department of Physics of Université de Montréal for a recruitment scholarship.

\bibliography{Arxiv_21-10-19}
\bibliographystyle{jpa}
\end{document}